\documentclass[sigconf, 12pt]{acmart}

\setcopyright{none}
\renewcommand\footnotetextcopyrightpermission[1]{}

\usepackage{balance}

\usepackage{makecell}
\usepackage{amsmath}
\usepackage{mathtools}
\usepackage{hyperref}
\usepackage{cleveref}
\usepackage[disable]{todonotes}

\newcommand{\job}[1]{\todo[color=teal!40,inline]{Bob for Joakim: #1}}
\usepackage{appendix}
\usepackage{listings}
\usepackage{booktabs}
\usepackage{csquotes}

\usepackage{afterpage}
\afterpage{\clearpage}

\makeatletter
\let\old@lstKV@SwitchCases\lstKV@SwitchCases
\def\lstKV@SwitchCases#1#2#3{}
\makeatother
\usepackage{lstlinebgrd}
\makeatletter
\let\lstKV@SwitchCases\old@lstKV@SwitchCases

\lst@Key{numbers}{none}{%
    \def\lst@PlaceNumber{\lst@linebgrd}%
    \lstKV@SwitchCases{#1}%
    {none:\\%
     left:\def\lst@PlaceNumber{\llap{\normalfont
                \lst@numberstyle{\thelstnumber}\kern\lst@numbersep}\lst@linebgrd}\\%
     right:\def\lst@PlaceNumber{\rlap{\normalfont
                \kern\linewidth \kern\lst@numbersep
                \lst@numberstyle{\thelstnumber}}\lst@linebgrd}%
    }{\PackageError{Listings}{Numbers #1 unknown}\@ehc}}
\makeatother

\lstdefinelanguage{diff}{
    basicstyle=\ttfamily\tiny,
    morecomment=[f][\color{gray}]{@@},
    morecomment=[f][\color{green}]{+\ },
    morecomment=[f][\color{orange}]{-\ },
}

\crefname{lstlisting}{listing}{listings}
\Crefname{lstlisting}{Listing}{Listings}

\lstdefinestyle{CStyle}{
    basicstyle=\small,
    breakatwhitespace=false,         
    breaklines=true,                 
    captionpos=b,                    
    keepspaces=true,                 
    numbers=left,                    
    numbersep=5pt,                  
    showspaces=false,                
    showstringspaces=false,
    showtabs=false,                  
    tabsize=4,
    language=C,
    frame=single
}

\newcommand*{\metaauthori}{Joakim Misund}
\newcommand*{\metaauthorii}{Bob Briscoe}
\newcommand*{\metashorttitle}{Disentangling Flaws in Linux DCTCP}
\newcommand*{\metatitle}{\metashorttitle}
\newcommand*{\metano}{TR-UIOJM-2022-004}
\newcommand*{\metakeywords}{Data Communication, Networks, Internet, Control, Congestion Control, TCP, Quality of Service, Performance, Responsiveness, Queuing Delay, Algorithm}
\newcommand*{\metainstitutioni}{University of Oslo}
\newcommand*{\metainstitutionii}{Independent}
\newcommand*{\metamaili}{joakimmi@ifi.uio.no}
\newcommand*{\metamailii}{research@bobbriscoe.net}
\newcommand*{\metaversion}{00B}
\newcommand*{\metadate}{18 Oct 2022}

\hypersetup{                       
	pdfauthor = {\metaauthori and \metaauthorii
	},
	pdftitle = {\metashorttitle},
	pdfsubject = {},
	pdfkeywords = {\metakeywords}
}%

\title{\metatitle}

\author{\metaauthori}
\affiliation{
	\institution{\metainstitutioni}
}
\email{\metamaili}

\author{\metaauthorii}
\affiliation{
	\institution{\metainstitutionii}
}
\email{\metamailii}
\date{}

\fancypagestyle{first}{%
	\fancyhead[LO,RE]{}%
	\fancyhead[LE,RO]{}%
	\fancyhead[C]{}%
}%

\begin{abstract}
	In the process of testing improvements to the Linux DCTCP code in various scenarios, we found different performance problems kept surfacing with no apparent pattern. This report records a systematic sequence of experiments designed to track down the causes of these problems, which were found to be due to a complex tangle of bugs and flaws. The report also provides and evaluates solutions in each case.
\end{abstract}

\begin{document}

\maketitle

\section{Introduction}
\label{sec:introduction}

\begin{figure}
	\centering
	\includegraphics[width=\linewidth]{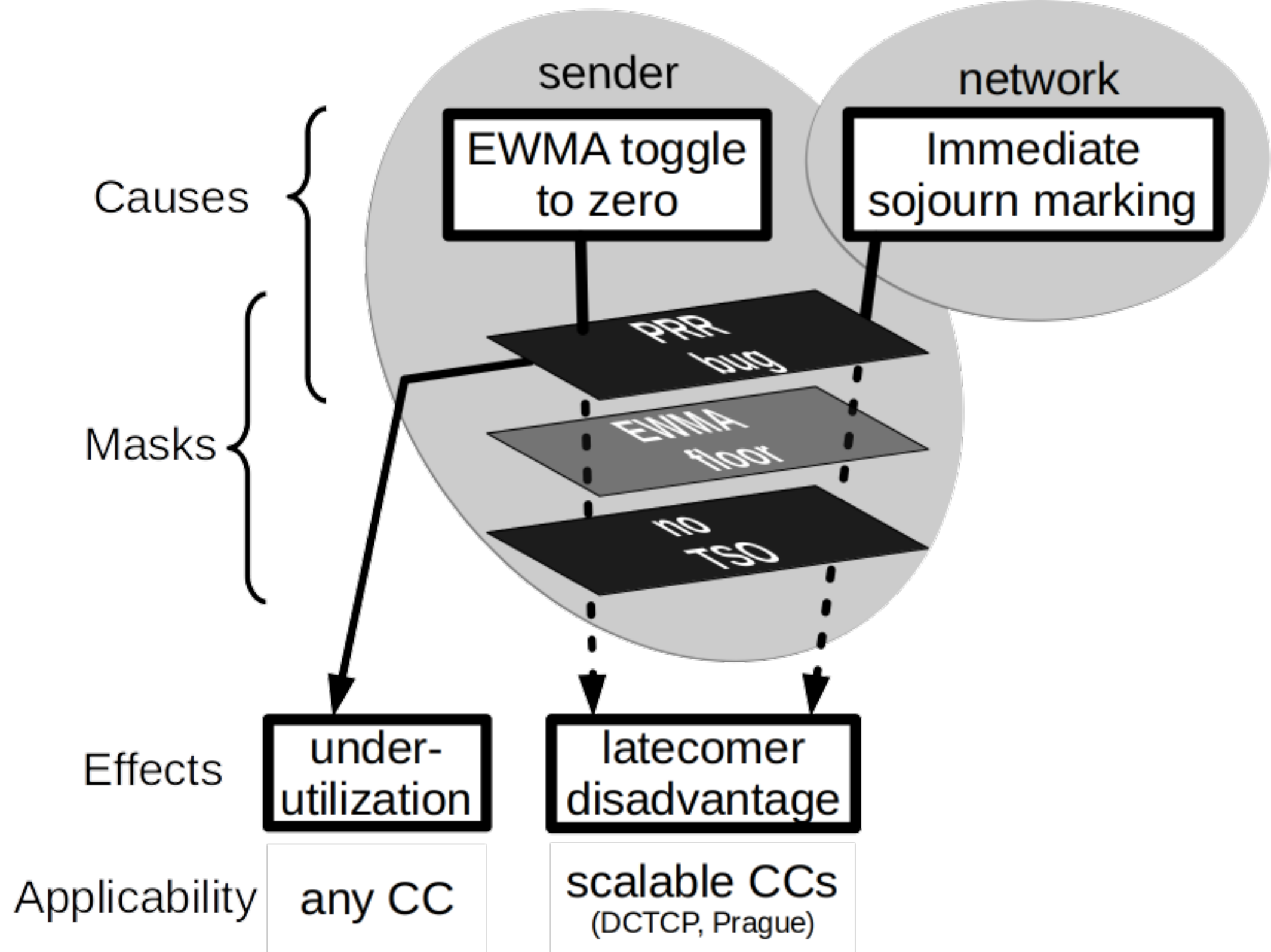}
	\caption{Simplified overview of the tangle of interactions between bugs \&
		flaws described in this paper.
		A bug in the Linux implementation of Proportional Rate Reduction causes
		underutilization, whatever the CCA (unless TSO is disabled ). When the 
		bug is patched, it unmasks a complex latecomer disadvantage problem, but 
		only for scalable CCAs such as DCTCP. To solve this, two underlying 
		causes both have to be fixed; i) A toggle to zero in the DCTCP sender's 
		congestion averaging; ii) use of sojourn time to measure instantaneous
		queue delay in the AQM. Just removing the toggle from DCTCP appears to mitigate
		the latecomer disadvantage. However, all it really does is set a floor to the
		EWMA, which partially masks the latecomer disadvantage (hence it is shown in
		grey), but it turns DCTCP into a classical CCA, losing the benefits of low delay
		variation and rapid tight congestion control.}
	\label{fig:paper_overview}
\end{figure}

This report records a sequence of experiments used to systematically track down the causes of a number of performance problems with the Linux implementation of DCTCP. It uncovers a tangle of interactions between subtle bugs and flaws in the Linux TCP stack, the DCTCP congestion control algorithm (CCA) and sojourn-based AQM marking, as shown in \cref{fig:paper_overview}. This introduction explains how all the bugs and flaws interrelate, with the benefit of hindsight. The report provides solutions for each bug or flaw, and experiments to measure the performance of each. A roadmap to the body of the report follows this introduction.

%
%

\paragraph{Proportion Rate Reduction (PRR)} is an experimental algorithm that attempts to
adjust the congestion window more gradually to the new ssthresh compared to Fast
Recovery. 
The claimed benefits are less time used in recovery and fewer timeouts~\cite{Dukkipati11:PRR}.

The Linux implementation of PRR is not as specified in RFC6937~\cite{rfc6937} due to a bug, which interacts with TCP Segmentation Offloading (TSO) to 
cause the congestion window to shrink more than intended. The consequences are lower
utilization due to each short stall being followed by a burst when the window corrects itself, which in turn can cause a congestion event to extend into another round of reduction.
\begin{quote}
\footnotesize Note: the Linux maintainers fixed the PRR bug after sight of an early draft of this report. Nonetheless, we still discuss the bug, to aid understanding of the other flaws it masked.
\end{quote}

The problem is most apparent when a congestion control algorithms (CCA) only reduces the window a
little in response to congestion, such as in Alternative Backoff with ECN (ABE~\cite{Khademi18:ABE}), and it is particularly apparent in
DCTCP~\cite{Alizadeh2010:DCTCP_Short}. 

In contrast, with the large congestion response of a traditional 
CCA, the troublesome code path
is avoided until towards the end of congestion window reduction (CWR). 
So, the bug does affect CUBIC, but less so than scalable CCAs like DCTCP.
Note that,
on a loss TSO is disabled, which lessens the impact of
this bug. 

\paragraph{Latecomer disadvantage:} Unfortunately, having fixed the PRR bug (by aligning the implementation with RFC 6937), we discovered it had been
masking another different and highly repeatable problem---a latecomer
disadvantage. This only affects scalable CCAs, not classical. 
When one long-running DCTCP flow arrives after another, the established flow maintains a competitive advantage---but only when TSO is enabled (and the PRR bug has been patched). It seems that 
the flows get stuck in a `local equilibrium' with unequal rates, that stops them converging further towards equality.

The latecomer disadvantage may not be such a concern in data centre networks for
several reasons. Firstly, as rate increases the burst size hits its upper size
limit. Secondly, NIC rates are usually more similar to core network link rates and so
often the NIC is the bottleneck, or the NIC is only a little faster than the bottleneck. Nonetheless, it is of concern for the use of scalable congestion controls in the Internet and other wide area networks.

We tracked down two causes of the latecomer disadvantage, one in the sender, one in the AQM; both of which have to be fixed to solve it:
\begin{itemize}
	\item A toggle to zero in the implementation of congestion averaging in the DCTCP CCA;
	\item Use of sojourn time to measure instantaneous queue delay in immediate AQMs used for DCTCP.
\end{itemize}

\paragraph{Toggle to Zero:}
Every round trip, DCTCP measures the ECN marking probability, and maintains a
moving average in a variable called alpha. The value of alpha is a small
fraction, and the larger the window, the smaller the value of alpha. Being
kernel code, DCTCP represents alpha with an integer (with resolution of 1 in
1024). However, it divides the change in alpha by 16 using a bit-shift, so alpha only moves when the relative change is greater than 16/1024. In Feb 2015 it was noticed that alpha could never reduce to zero. So rather quirky code was added to toggle values smaller than 16 down to zero~\cite{shewmaker15:Linux_DCTCP_EWMA}, without any mention of performance testing results. Also a truncation problem was left, where the fraction of ECN feedback in each round-trip was rounded down to the nearest 1/16, which always black-holed any feedback lower than 1/16.

The toggle to zero does often mask the latecomer disadvantage, probably by jolting the flows out of their local equilibrium so that they converge to equal rates.
However, it doesn't work in all scenarios, and it causes more queue variation than expected for DCTCP. 

Removing the toggle to zero also appears to partially resolve the latecomer disadvantage. However, further investigation found that this was only because it leaves a floor to the EWMA at 16/1024, which effectively turns DCTCP into a classical CCA with a fixed floor to its multiplicative window decrease. Again, this is often enough to jolt the flows out of their local equilibrium. However, it loses the advantage of scalability. As flow rate scales, the time between congestion signals grows. Then, for instance, when capacity becomes available, congestion signalling will go quiet, but it takes a prohibitively long time to detect that it is a real absence of congestion when the normal recovery time is so long, like a classical CCA. So there is no potential to modify the CCA to fill available capacity rapidly, which there should be with a scalable CCA.

The implementation of a derivative of DCTCP called Prague~\cite{Briscoe19a:TCP_Prague_Linux, Briscoe21b:PragueCC-ID} fixes both these problems (the toggle and the floor) by using a more precise (upscaled) EWMA variable. However, with just two long running flows and no background traffic, the latecomer disadvantage becomes more prevalent again, for the reasons tracked down below. Briefly, the root cause of the latecomer disadvantage is in the AQM so, unless the AQM is fixed, a more precise EWMA no longer masks the problem, because it no longer jolts the flows out of their local equilibrium.

\paragraph{Use of sojourn time in the AQM}
Superficially, the latecomer disadvantage seems to be due to the way Linux
limits the size of TSO bursts so that (by default) they will cause no more than
1\,ms of queue delay at the bottleneck. The calculation in Linux results in a
burst size (in packets) that is  proportional to a flow's congestion window
(cwnd). Therefore, when a new flow starts from a small cwnd, Linux keeps its
bursts small, but it allows pre-existing flows to produce larger
bursts, because they have a larger cwnd.

This alone should not cause the latecomer disadvantage. However, with scalable CCAs like DCTCP and Prague, we found
that a flow sending packets in larger bursts induces a \emph{lower} ECN-marking
probability at the bottleneck AQM. The new entrant with smaller bursts then sees
proportionately higher congestion, so it can never catch up with the flow that
has established a higher cwnd and therefore sends in larger bursts, which the AQM perversely marks less.

In work being published separately, this perverse ECN marking was found to be because the AQM applies ECN marking to packets with greater queue delay as measured by their own sojourn time. This biases marking onto those packets that suffer most delay \emph{themselves}, which tend to be those behind a burst. We have implemented an alternative marking scheme that focuses marking onto those packets that cause queuing to \emph{other} packets behind them. This focuses more marking on the packets in the bursts, and hardly any on those from smoother flows caught behind them. The experiments in this report show that this corrects the marking perversity, and resolves the latecomer disadvantage.\footnote{However, with bursts greater than the marking threshold it would be hard for any marking scheme to save smooth traffic from being marked. So some form of burst policing would probably also be necessary.}

\subsection{Roadmap}
\label{roadmap}

\Cref{sec:methodology} explains the systematic approach taken to tracking down the causes of the various performance problems, and tabulates the notation we use for the different code variants. Then each main section of the report works through the three main causes of performance problems:
\begin{description}
	\item[\Cref{sec:prr}:] The PRR bug and solution;
	\item[\Cref{sec:late-comer}:] The latecomer disadvantage;
	\item[\Cref{sec:toggling-issue}:] Improving the DCTCP EWMA as a potential solution to the latecomer disadvantage. Including removing the toggle to zero and investigating why fixing the truncation and rounding bugs makes the latecomer disadvantage slightly worse than just setting a floor to the EWMA value;
	\item[\Cref{sec:altering-AQM}:] The interaction between segmentation offload at the sender and the blame-shifting problem in AQMs using sojourn-based marking (problems and solution).
\end{description}

\Cref{sec:tail-pieces} gives conclusions, recommendations and proposed further work, followed by additional information in appendices:
\begin{description}
	\item[\Cref{appendix:dctcp-algo-and-implementation}:] Tutorial material on the design and implementation of DCTCP in Linux; 
	\item[\Cref{appendix:machine-config}:] Full specification of the experiments to support independent verification;
	\item[\Cref{appendix:code_variants}:] Links to diffs between the DCTCP code and the variant  in each experiment.
\end{description}

\clearpage
\onecolumn
\section{Methodology}
\label{sec:methodology}
To investigate potential issues we implemented multiple variants of DCTCP and tested
them in different scenarios. The two testbeds used and
their configurations are described in \cref{appendix:machine-config}.
The DCTCP algorithm and key parts of the Linux implementation are described in
\cref{appendix:dctcp-algo-and-implementation}.
We decided to use DCTCP throughout this paper because it suffers from all of the
problems described. Wherever a problem affects other CCAs
such as TCP Cubic and TCP Prague they are also included. 

\subsection{DCTCP variants}
\label{sec:dctcp-variants}

\begin{table}
   \begin{tabular}{c l}
   	Code & Meaning\\
   	\hline
	P & PRR enabled \\
	S & TSO enabled \\
	nn & 10 or 20-bit precision of EWMA (dctcp\_alpha)\\
	T & EWMA toggles to zero below \(16/2^{\mathrm{nn}}\)\\
	U & EWMA stored upscaled for precision (see \cref{sec:toggling-solution})\\
	\\
  \end{tabular}
  \caption{Codes for DCTCP variants tested.  Whether a letter is in capitals or lower case respectively indicates whether the capability is used or not.}
  \label{tab:dctcp-codes}
\end{table}

\begin{table}
	\begin{tabular}{r p{5in}}
		Code & Description\\
		\hline
		DCTCP-PS10Tu & Linux DCTCP default (Sep'22)\\
		DCTCP-PS10tu & Linux pre v4.3 DCTCP default (pre-2-Nov-2015) \\
        DCTCP-pS20tU & Linux DCTCP configured as equivalent to TCP Prague default (Sep'22)\\
        \\
	\end{tabular}
	\caption{Example variant codes that represent particular releases.}
	\label{tab:dctcp-ex-variants}
\end{table}

We use the five different 2-state codes in \cref{tab:dctcp-codes} to specify which variant of DCTCP we are talking about at any one time.\footnote{Early in the report, P means the bugged PRR code. But once we have described how to fix it, in \S\,\ref{sec:late-comer} onwards, P means with PRR fixed and enabled. Also, in \S\,\ref{sec:altering-AQM} an additional configuration option is introduced in TCP Prague for the max burst (1\,ms, 250\,\(\mu\)s or no burst).} Some example variants are listed in \cref{tab:dctcp-ex-variants}, with the particular release they represent. \Cref{sec:dctcp-diffs} lists all the variants used in this report, with links to online diffs of each relative to the current DCTCP implementation.

PRR is circumvented by implementing the cong\_control callback 
introduced with BBR. This is accompanied by updating the pacing rate and
congestion reaction in the DCTCP module.

\subsection{AQM variants}
\label{sec:aqm}
We have used an AQM that can mark packets using a threshold or a ramp based on
instantaneous sojourn time. Various threshold values are configured, as stated in each experiment. Sojourn time is the time a packet has spent in the
queue, i.e.\ the elapsed time from enqueue to dequeue.

In \S\,\ref{sec:altering-AQM} an additional AQM configuration option is introduced to compare marking based on sojourn time with that based on expected service time (EST), which is the time it is expected to take to drain the backlog behind the head packet, assuming the recent drain rate continues.

\clearpage
\section{Bug and issue in PRR}
\label{sec:prr}

\subsection{Bug: Unintentional cwnd decrease during CWR}
\label{sec:prr-bug}

Proportion Rate Reduction (PRR) is an experimental algorithm that attempts to
adjust the congestion window more gradually to the new ssthresh compared to Fast
Recovery. PRR accounts for the number of packets that have been
acked and sent throughout a congestion event and uses this with the new
ssthresh to dynamically adjust the number of new packets that can be sent. By
the end of recovery the congestion window will be close to the new ssthresh.
The claimed benefits are less time used in recovery and less timeouts~\cite{Dukkipati11:PRR}.

An interaction between a bug in Linux PRR and TCP Segmentation Offloading (TSO) causes under-utilization and possibly more bursty traffic.
When combined with TCP Segmentation Offloading (TSO), 
the bug causes the congestion window to shrink more than intended. When TSO 
stores up segments to send all in one
go, the bugged PRR code counts them as if they have already been transmitted.
The consequence is lower utilization and possibly more bursty traffic. 

The problem is most apparent when a congestion control only reduces the window a
little in response to congestion. It is therefore most problematic in recent
congestion control algorithms (CCAs) that respond less to ECN markings than to loss, for instance
those that use the Alternative Backoff with ECN (ABE~\cite{Khademi18:ABE})
modification. And it is particularly apparent in
DCTCP~\cite{Alizadeh2010:DCTCP_Short}.

In contrast, with the large congestion response of a traditional 
CCA the troublesome code path
is avoided until towards the end of congestion window reduction (CWR). 
So, the bug does affect CUBIC, but less so than scalable CCAs like DCTCP.
Note that,
if congestion is caused by loss, then TSO is disabled which lessens the impact of
this bug. 

The bug is caused by a difference between the implementation and the
specification in RFC6937~\cite{rfc6937}. 
The TSO code defers using the allowance given in the PRR 
code, and this allowance is overwritten with a new allowance on the reception of
the next acknowledgement (ack). Previously given allowance is assumed to be
used, but this is not the case when TSO defers sending. We will go into more detail in
\cref{sec:prr-code-inspection} after we have looked at some illustrations of the problem.

\paragraph{PRR bug: Data centre network scenario}
\begin{figure}
  \centering
  \includegraphics[width=.8\linewidth]{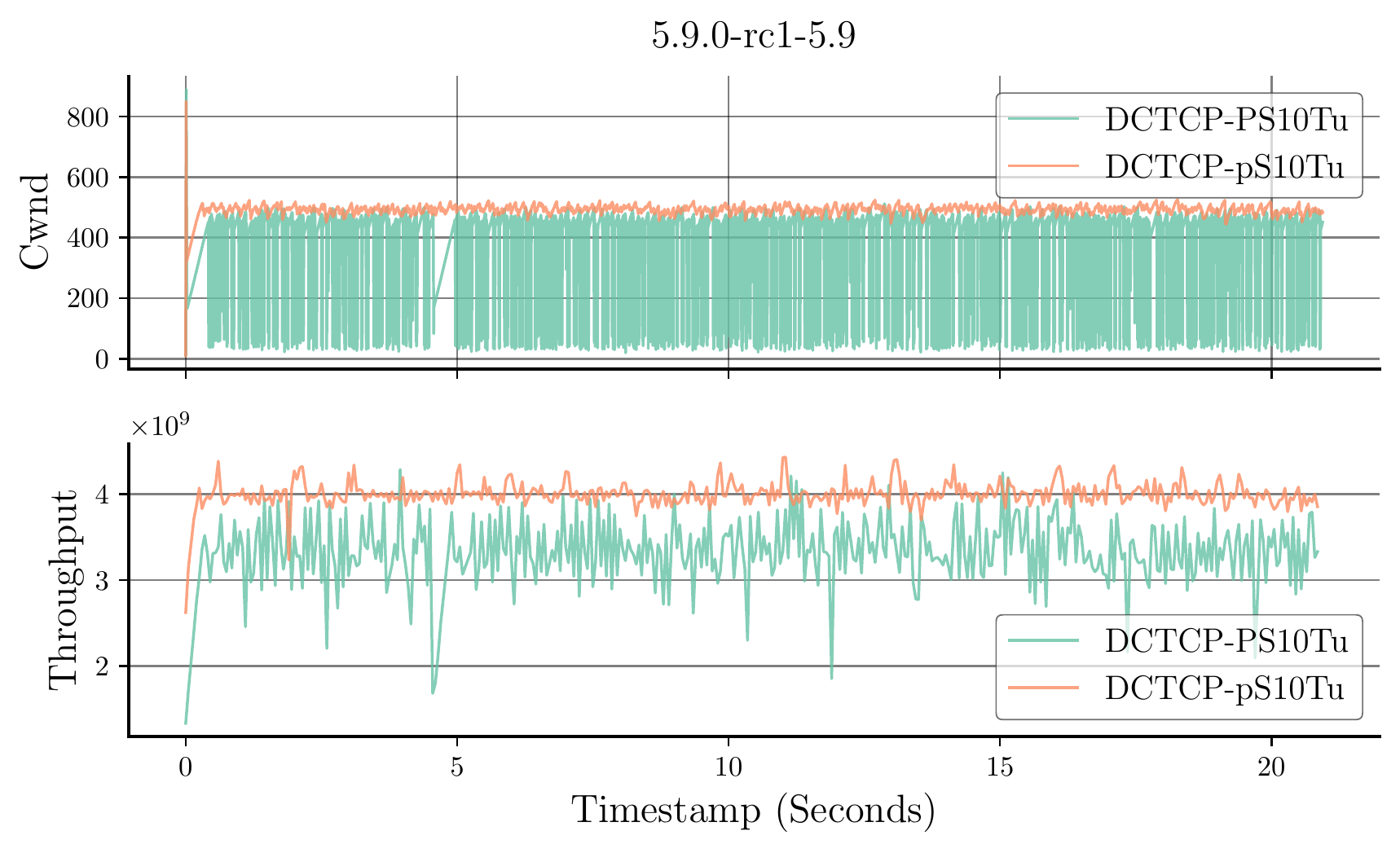}
  \caption{The Effect of the PRR bug in a Data Centre scenario.\\
  	DCTCP with PRR (DCTCP-PS10Tu) suffers from lower utilization because it almost halts
    sending during recovery and it ends up sending larger bursts when compared with
    DCTCP without PRR (DCTCP-pS10Tu).\\
    Capacity: 5\,Gb/s; RTT: 1\,ms; AQM: step at 163\,us.}
  \label{fig:1ms5000mbps-issue-illustration}
\end{figure}

The issue is noticeable at data centre rates as can be seen in
\cref{fig:1ms5000mbps-issue-illustration}. The link rate is 5\,Gb/s%
\footnote{Even without PRR, the data rate has trouble exceeding
	4\,Gb/s. There are two plausible explanations: i) to enforce a rate of 5\,Gb/s
	we use htb, which we discovered was mis-configured to have no burst
	allowance, meaning that idle time goes to waste; ii) the step used was too low and allowed system noise to prematurely trigger
	CE-marks.\job{We cannot publish this report with this as the only evidence of the problem in a DC. I'm afraid we need to re-run this experiment properly so we can lose the footnote. Initial investigations: 600B burst allowance on htb.}} 
and the
round-trip time is 1ms. The AQM is a step at 163\,us. The different variants are run
in separate experiments. The only difference between the two DCTCP variants is
the use of PRR. DCTCP-PS10Tu is DCTCP with PRR (regular DCTCP), and DCTCP-pS10Tu is DCTCP
without PRR.

DCTCP with PRR suffers from lower throughput due to sending being almost halted
on ECN events and increased burstiness when the congestion window is
recovered after ECN events. At
almost every congestion event cwnd decays to a very small value before it jumps
back up again. This jump seem to cause a large burst of packets to be sent and a
new congestion event.

\paragraph{PRR bug: Wide area network scenario}
\begin{figure}
  \centering
  \includegraphics[width=.8\linewidth]{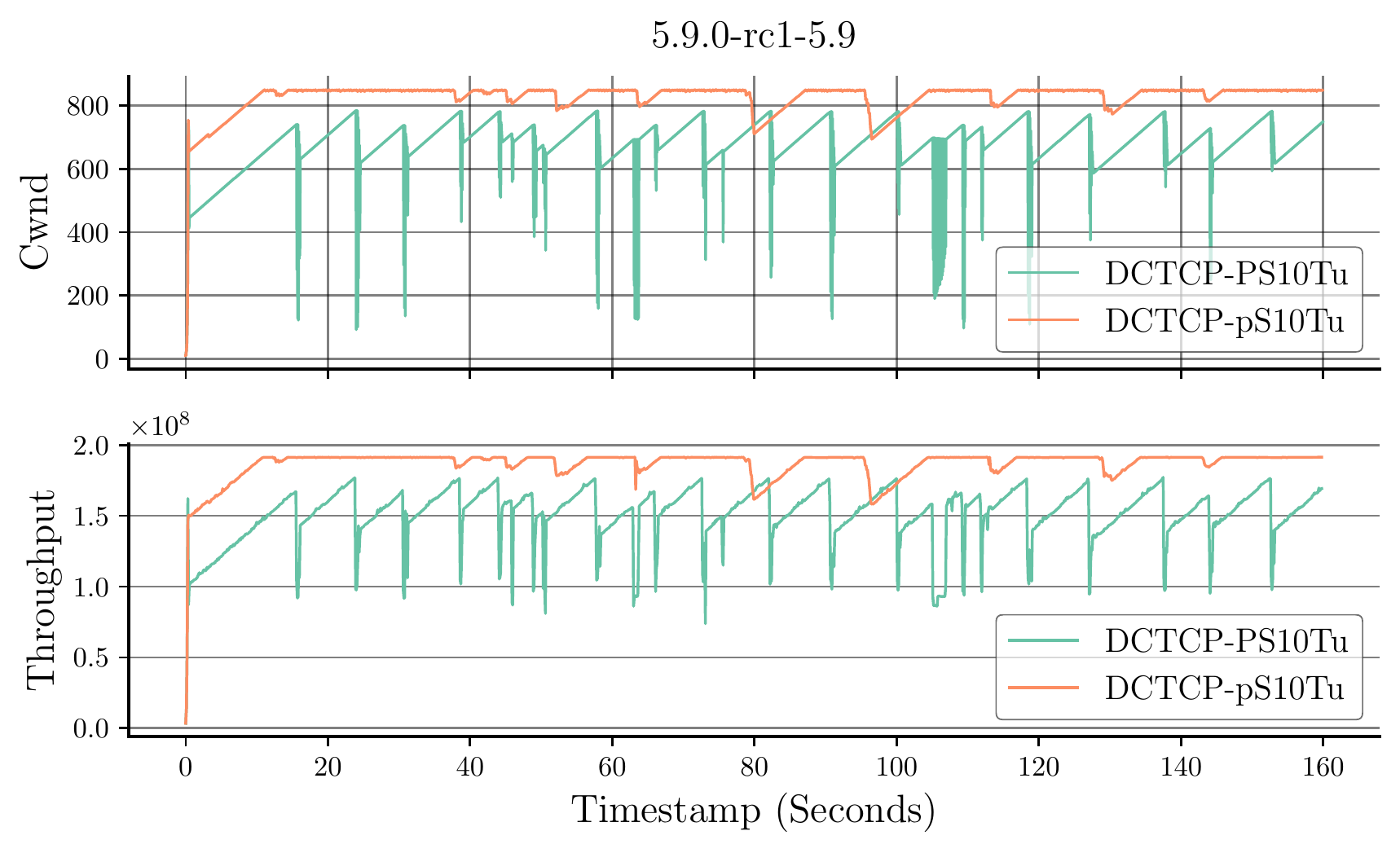}
  \caption{The Effect of the PRR bug in a Wide Area Network scenario 
  	is similar to that in the Data Centre scenario of \Cref{fig:1ms5000mbps-issue-illustration}\\
    Capacity: 200\,Mb/s; RTT: 50ms; AQM: step at 1\,ms.}
  \label{fig:50ms200mbps-issue-illustration}
\end{figure}

The issue is also present in wide are network scenarios as we show in 
\cref{fig:50ms200mbps-issue-illustration}. The figure shows the
evolution of cwnd and throughput for a single flow using regular DCTCP and DCTCP
without PRR.

\begin{figure}
  \centering
  \includegraphics[width=.8\linewidth]{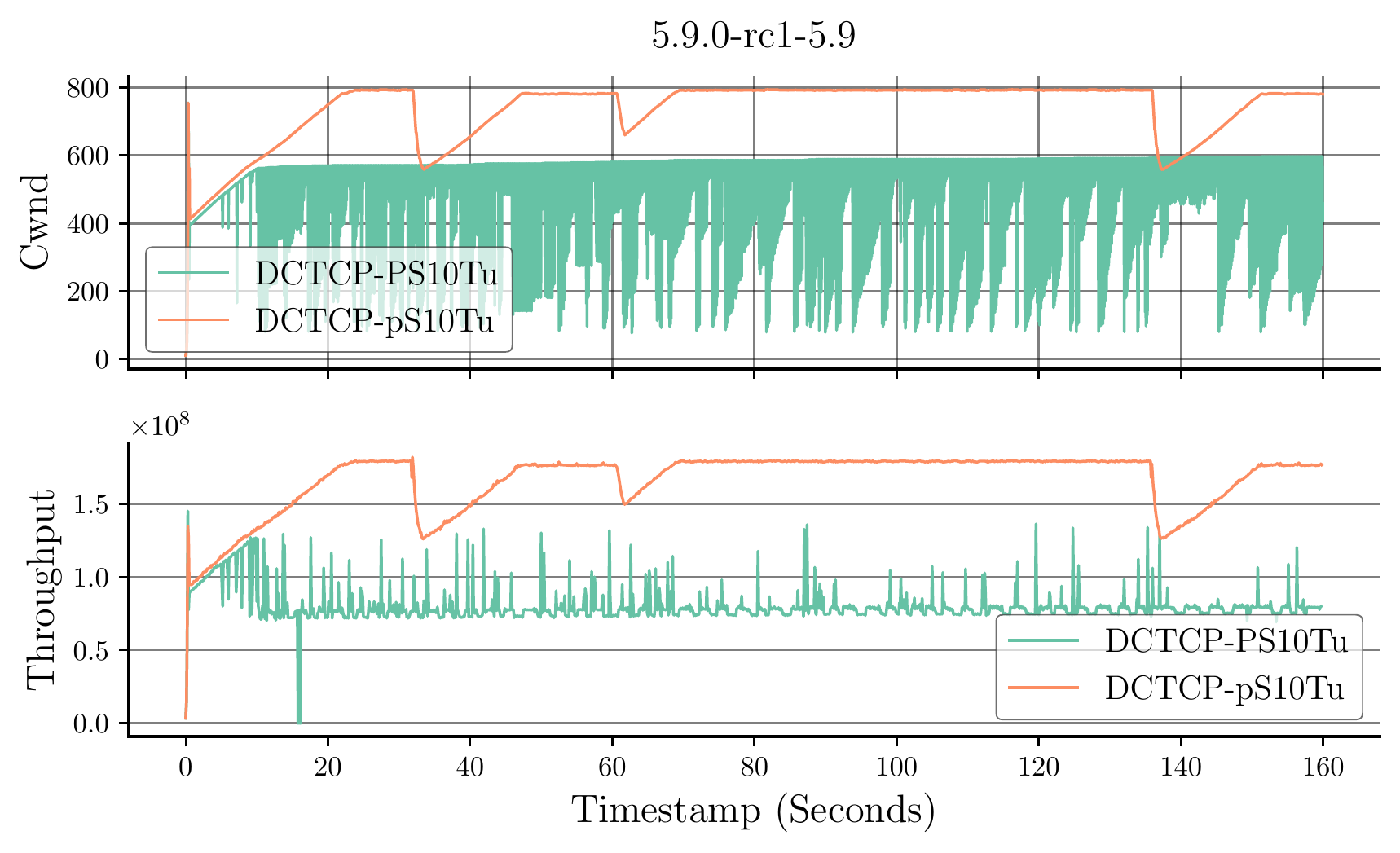}
  \caption{Inducing more frequent congestion events by using a ramp AQM makes the consequence of the
    bug more visible.\\
    Capacity: 200\,Mb/s; RTT: 50\,ms; AQM: ramp from 0.5\,ms to
    2.0\,ms instead of a step threshold.}
  \label{fig:50ms200mbps-issue-illustration-ramp}
\end{figure}

The bug in PRR is even clearer if a ramp from 0.5\,ms to 2.0\,ms is used to mark
packets. This can be seen in
\cref{fig:50ms200mbps-issue-illustration-ramp}. The reason is that
congestion events are much more frequent because cwnd oscillation is smaller.

\begin{figure}
  \centering
  \includegraphics[width=.8\linewidth]{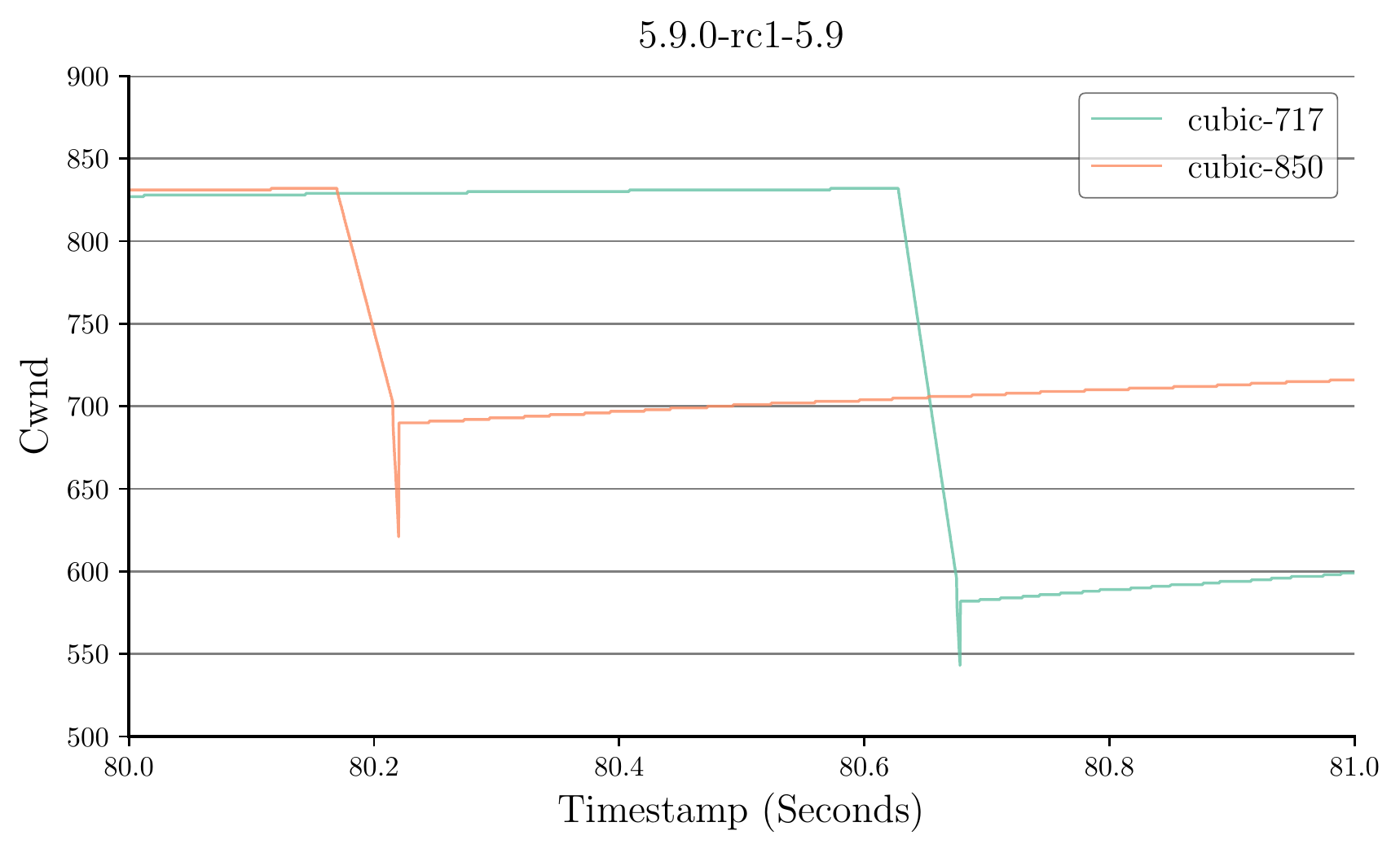}
  \caption{Excerpt of cwnd time sequence plot for two single-flow experiment runs using
    TCP ECN-CUBIC with beta parameter 717 (the default of 0.7) and 850 (0.83). The PRR bug causes the downward spikes at the bottom of each sawtooth, the main impact of
    which is a more bursty sending pattern when the congestion window
    recovers.\\
    Capacity: 200\,Mb/s; RTT: 48\,ms; AQM: step at 2\,ms.}
  \label{fig:48ms200mbps-issue-illustration-cubic}
\end{figure}

The bug is also noticeable when TCP Cubic is used.%
\footnote{Here, a DCTCP-like
	AQM is used, with immediate marking at a shallow threshold. This is not a common
	configuration for CUBIC. With a deeper buffer, the bug might not greatly affect
	CUBIC throughput. However, the configuration in this experiment is similar to a 
	shallow tail-drop buffer, where PRR undershooting would affect throughput largely 
	as shown.} 
The default value for the
back-off factor (beta) in Linux is 717 (0.7), but RFC 8511 suggest that increasing it
to 0.85 can provide benefits. However, increasing beta makes the PRR bug more
noticeable as can be seen in
\cref{fig:48ms200mbps-issue-illustration-cubic}. It is possible that there is little
impact on utilization in scenarios with sufficient queueing during the
slight halt in sending. There is however a negative impact on temporary queueing
delay because the sending pattern is more bursty when the congestion window recovers.

\clearpage

\begin{minipage}{\linewidth}
\begin{lstlisting}[
  caption={PRR pseudocode in RFC6937},
  label={pseudocode:PRR},
  basicstyle=\footnotesize,
  firstnumber=298,
  firstline=298,
  lastline=315,
  frame=single,
  numbers=left,
  breaklines=True,
  linebackgroundcolor={
    \ifnum \value{lstnumber} = 314 \color{orange} \fi \ifnum \value{lstnumber} = 309 \color{orange} \fi}]
   On every ACK during recovery compute:

      DeliveredData = change_in(snd.una) + change_in(SACKd)
      prr_delivered += DeliveredData
      pipe = (RFC 6675 pipe algorithm)
      if (pipe > ssthresh) {
         // Proportional Rate Reduction
         sndcnt = CEIL(prr_delivered * ssthresh / RecoverFS) - prr_out
      } else {
         // Two versions of the Reduction Bound
         if (conservative) {    // PRR-CRB
           limit = prr_delivered - prr_out
         } else {               // PRR-SSRB
           limit = MAX(prr_delivered - prr_out, DeliveredData) + MSS
         }
         // Attempt to catch up, as permitted by limit
         sndcnt = MIN(ssthresh - pipe, limit)
      }
\end{lstlisting}

\begin{lstlisting}[
  caption={PRR implementation in Linux},
  label={code:PRR},
  firstline=2479,
  firstnumber=2479,
  lastline=2504,
  breaklines=True,
  style=cstyle,
  basicstyle=\footnotesize,
  linebackgroundcolor={
    \ifnum \value{lstnumber} = 2503 \color{orange} \fi \ifnum \value{lstnumber} = 2499 \color{orange} \fi}]
void tcp_cwnd_reduction(struct sock *sk, int newly_acked_sacked, int flag)
{
	struct tcp_sock *tp = tcp_sk(sk);
	int sndcnt = 0;
	int delta = tp->snd_ssthresh - tcp_packets_in_flight(tp);

	if (newly_acked_sacked <= 0 || WARN_ON_ONCE(!tp->prior_cwnd))
		return;

	tp->prr_delivered += newly_acked_sacked;
	if (delta < 0) {
		u64 dividend = (u64)tp->snd_ssthresh * tp->prr_delivered +
			       tp->prior_cwnd - 1;
		sndcnt = div_u64(dividend, tp->prior_cwnd) - tp->prr_out;
	} else if ((flag & (FLAG_RETRANS_DATA_ACKED | FLAG_LOST_RETRANS)) ==
		   FLAG_RETRANS_DATA_ACKED) {
		sndcnt = min_t(int, delta,
			       max_t(int, tp->prr_delivered - tp->prr_out,
				     newly_acked_sacked) + 1);
	} else {
		sndcnt = min(delta, newly_acked_sacked);
	}
	/* Force a fast retransmit upon entering fast recovery */
	sndcnt = max(sndcnt, (tp->prr_out ? 0 : 1));
	tp->snd_cwnd = tcp_packets_in_flight(tp) + sndcnt;
}
\end{lstlisting}
\end{minipage}

\subsubsection{Differences between RFC6937 and PRR implementation}
\label{sec:prr-code-inspection}
PRR is described in RFC6937. It includes pseudo-code relisted in
\Cref{pseudocode:PRR}. The 'On every ACK' part of the implementation in Linux is
listed in \cref{code:PRR}.

In the draft it says the following:
\blockquote {We introduce a local variable "sndcnt",
  which indicates exactly how many bytes should be sent in response to each
  ACK.}

That means that each time a new ack is received the cwnd should be increased
enough so that sndcnt bytes are sent. Linux operates in number of packets instead
of number of bytes; so we will continue the explanation in terms of number of
packets. The Linux implementation executes the increase in line 2503.
It adds sndcnt to the current number of packets in
flight. However, it never checks that those sndcnt packets are sent, so if they
aren't, the next update will simply discard the previous increase. There is a difference
between allowing sndcnt more packet to be sent and actually sending them. If the
opportunity is not seized, the allowance is thrown away immediately on the reception of the
next acknowledgement. The crux is that packets that should be sent according to
the rules of PRR are not sent because TSO defers sending in an attempt to create
larger bursts.

There is a difference between the implementation of PRR in Linux and the
pseudocode in RFC6937 in the clause called PRR-CRB. We will use the variable
names in Linux for the following discussion. The variable delta is
the difference between ssthresh and pipe (in-flight packets). We define the
allowance surplus to denote the difference between prr\_delivered and prr\_out.

In the pseudocode, the limit (the number of packets that can be sent in
response to an ack) is set to the maximum of delta and the allowance
surplus. This makes sure that cwnd is recovered if it drops below ssthresh.
However, on line 2499 in the implementation of PRR, Linux uses the minimum of the
number of newly delivered segments and delta. Consequently, if cwnd falls
more than the number of newly delivered segments below ssthresh, it will not
recover before the end of CWR. This can be fixed by aligning Linux's
implementation with the pseudocode in RFC6937.

\subsubsection{Fix: Aligning Linux's PRR implementation with RFC6937}
\label{sec:prr-fix-align-rfc}

As we discussed in \ref{sec:prr-code-inspection} the behaviour is a result of a
difference in the pseudocode in RFC6937 and the implementation in Linux.
The implementation in linux misbehaves
when the number of in-flight packets is less than the target ssthresh.

It is possible that the algorithm was intentionally changed in fear of
introducing bursty behaviour, but the other conditions can also introduce bursty
behaviour, therefore this seems unlikely.

\begin{minipage}{\linewidth}
\begin{lstlisting}[
  caption={Patch applied to align PRR with RFC6937},
  label={Patch:PRR},
  basicstyle=\footnotesize,
  frame=single,
  numbers=left,
  breaklines=True]
diff --git a/net/ipv4/tcp_input.c b/net/ipv4/tcp_input.c
index 184ea556f50e..556dec987251 100644
--- a/net/ipv4/tcp_input.c
+++ b/net/ipv4/tcp_input.c
@@ -2502,7 +2502,7 @@ void tcp_cwnd_reduction(struct sock *sk, int newly_acked_sacked, int flag)
 			       max_t(int, tp->prr_delivered - tp->prr_out,
 				     newly_acked_sacked) + 1);
 	} else {
-		sndcnt = min(delta, newly_acked_sacked);
+		sndcnt = min(delta, tp->prr_delivered - tp->prr_out);
 	}
 	/* Force a fast retransmit upon entering fast recovery */
 	sndcnt = max(sndcnt, (tp->prr_out ? 0 : 1));
\end{lstlisting}
\end{minipage}

The patch we propose is listed in \cref{Patch:PRR}. It takes the difference
between what it should have been sent and what has been sent into account
when it calculates the new allowance.

\paragraph{PRR bugfix: Data center network scenario}
\begin{figure}
  \centering
  \includegraphics[width=.8\linewidth]{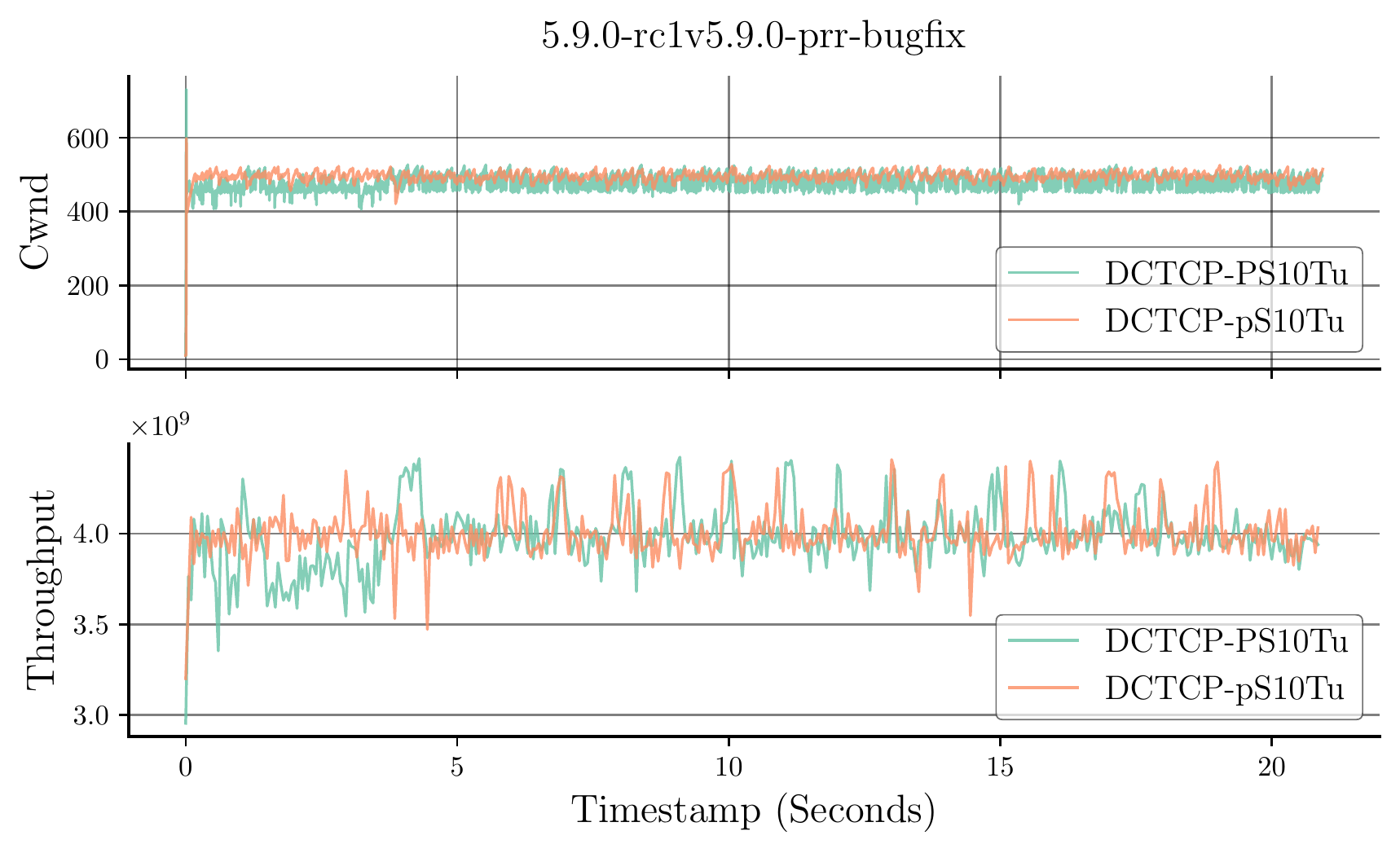}
  \caption{Fixed PRR bug in a data centre network scenario (c.f.\ \Cref{fig:1ms5000mbps-issue-illustration}).\\
  	Aligning Linux's implementation with RFC6937 fixes the PRR bug and its
    consequences.\\
    Capacity: 5\,Gb/s; RTT: 1\,ms; AQM: step at 1\,ms.}
  \label{fig:1ms5000mbps-RFC-aligned-illustration}
\end{figure}

\Cref{fig:1ms5000mbps-RFC-aligned-illustration} shows that aligning Linux's
implementation with the RFC prevents the cwnd from decreasing rapidly during
CWR. The cwnd is still reduced by a fraction of the burst size because packets
waiting in the TSO defer state are left out in the pipe-calculation upon
entering CWR. The number of in-flight packets is less than the
congestion window at the start of CWR because of TSO deferral.

\paragraph{PRR bugfix: Wide area network scenario}
\begin{figure}
  \includegraphics[width=.8\linewidth]{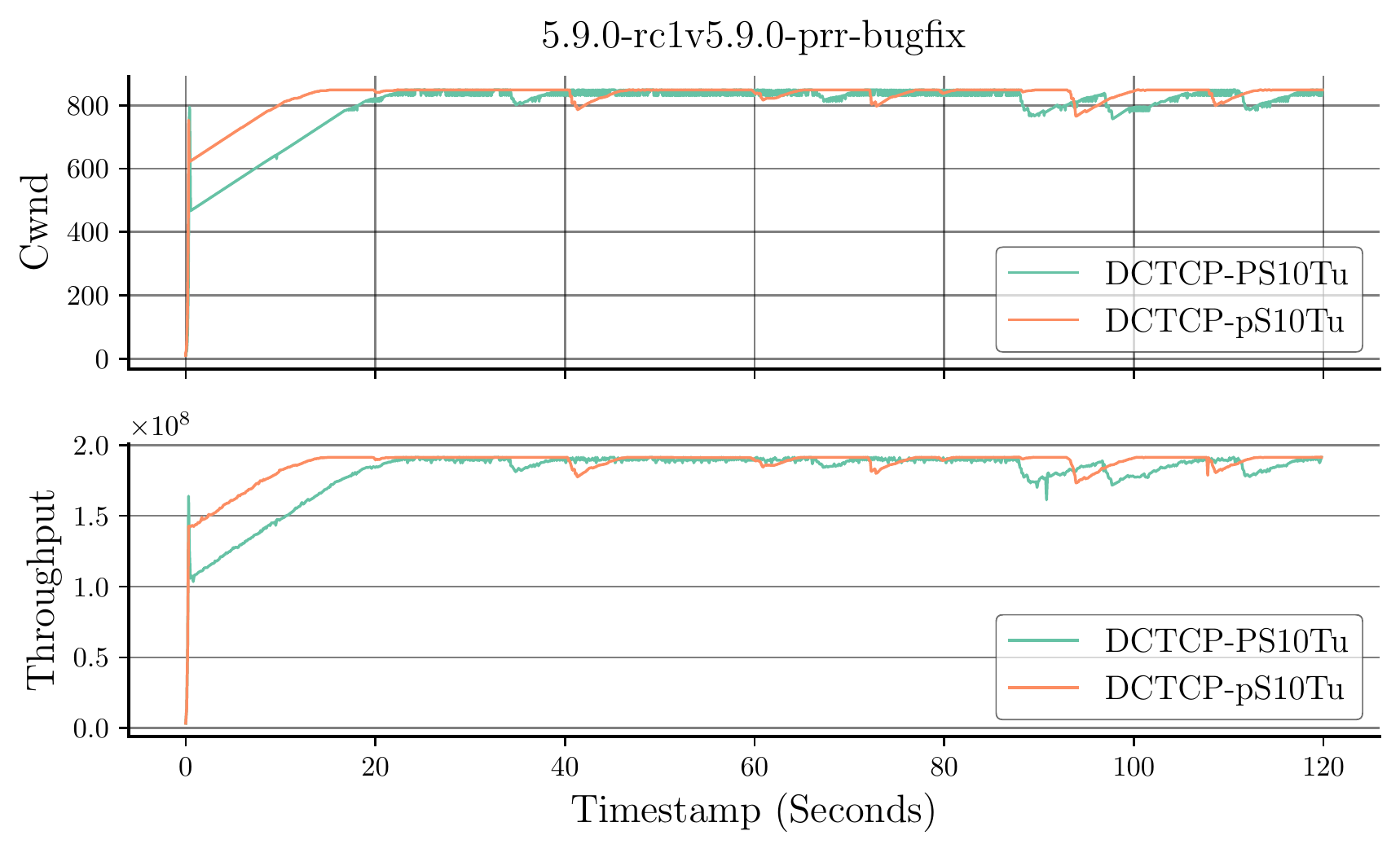}
  \caption{Fixed PRR Bug in a wide area network scenario (c.f.\ \Cref{fig:50ms200mbps-issue-illustration}).\\
  	Aligning Linux's implementation with RFC6937 fixes the PRR bug and its
    consequences.\\
    Capacity: 200\,Mb/s; RTT: 50\,ms; AQM: step at 1\,ms.}
  \label{fig:50ms200mbps-RFC-aligned-illustration}
\end{figure}
\begin{figure}
  \centering
  \includegraphics[width=.8\linewidth]{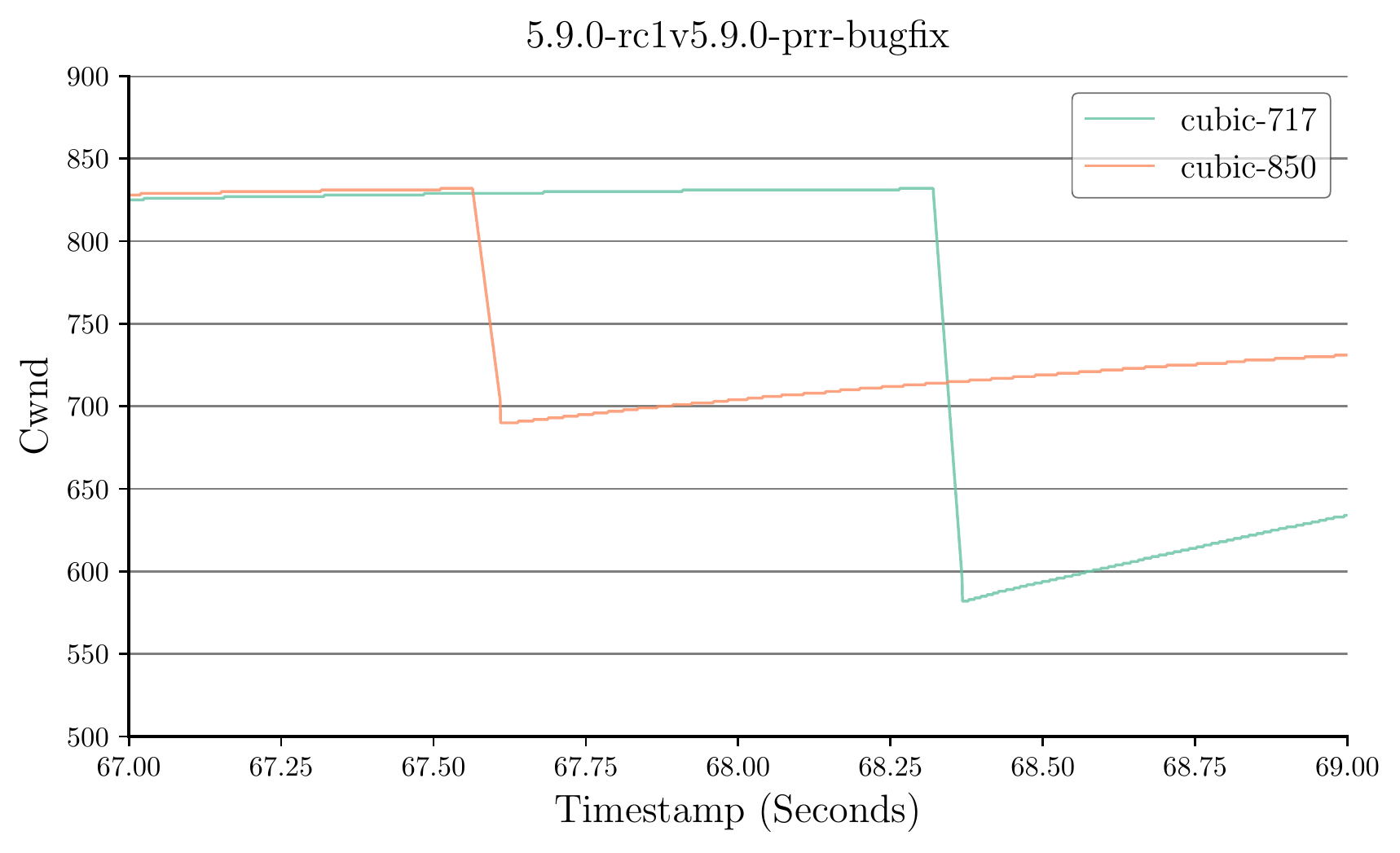}
  \caption{TCP CUBIC congestion window behaviour is also fixed (c.f.\ \Cref{fig:48ms200mbps-issue-illustration-cubic}).\\
  	Excerpt of cwnd time sequence plot for two single flows using
    TCP CUBIC with beta parameter 717 (0.7) and 850 (0.83).\\
    Capacity: 200\,Mb/s; RTT: 48\,ms; AQM: step at 2\,ms.}
  \label{fig:48ms200mbps-RFC-aligned-illustration-cubic}
\end{figure}

\Cref{fig:50ms200mbps-RFC-aligned-illustration,fig:48ms200mbps-RFC-aligned-illustration-cubic}
shows that the bugfix removes the issue in a wide are network scenario whether using DCTCP or
TCP CUBIC.

The PRR bug affects data centre and wide area network scenarios. It might affect
data centres more because the burst sizes at higher rates are greater, and so
TSO is more unlikely to allow a small number of packets to be sent.

\subsection{PRR bug: Discussion}
It is unknown to what degree the PRR bug has affected researchers' work over the
last couple of years. It is likely that utilization and fairness changes
after applying this bugfix. Every experiment that has used the default DCTCP
implementation might need revision. For instance, two recent interesting papers on L4S use
DCTCP in their experiments~\cite{BoruOljira:L4S_validate,Nadas20:PPV_L4S_Scheduler}.
They do not mention that PRR was disabled, so it is possible that the bug impacted the results.

It is unlikely that production deployments of DCTCP do not suffer from this bug. 
It is hard for us to tell whether or not this bug causes significant
loss of efficiency in production environments. 
It is therefore important to undertake proper comparisons between the bugged and fixed code 
with realistic traffic.


\subsection{PRR \& TCP Prague}
\label{sec:prr:discussion-tcp_prague}

The implementation of TCP Prague, created and maintained by the L4S team, avoids
PRR altogether by implementing the cong\_control callback introduced with
BRR. It is therefore not affected by the PRR bug at all.

Interestingly, TCP Prague has a similar gradual reduction to handle sub-MSS
adjustments to it's congestion window. The implementation, however, does not
spread the adjustment over a whole round-trip time. It makes at most an
adjustment of 1\,MSS per ack. That way it prevents sudden bursts caused by large
changes in its congestion window.

\subsection{To PRR, or not to PRR, with Scalable CCAs}
\label{sec:prr-issue}
The consequence of using the number of in-flight packets to build a burst when TSO
defers sending is that the number of in-flight packets can be
less than the congestion window when CWR is entered. If the change in ssthresh
is small this can cause the congestion window to start
below the new ssthresh, and stay so throughout the whole round-trip time. The
problem is barely noticeable for Classic CCAs because their
reductions are usually larger than the burst size TSO tries to achieve. RFC~6937
does not discuss the interaction between PRR and TSO deferral, so it is unclear
to us whether this is intended behaviour or not.

DCTCP is a scalable CCA that tries to adjust its congestion
window gradually. PRR prevents it from doing so because it forces a larger
reduction if there is a greater number of packets waiting to be sent as a TSO
burst when CWR is entered compared to DCTCP's calculated reduction.
If congestion events are very frequent, PRR forces DCTCP to maintain a lower
window than it intends.

The fact that the congestion window can stay below the new ssthresh for the
whole duration of CWR is problematic for two reasons.
\begin{enumerate}
	\item It reduces the load on and possibly the utilization of the bottleneck
	link. A very slight reduction in ssthresh that would fully utilize the link can
	be overruled by a larger reduction decided by PRR.
	\item It might confuse the CCA. If an algorithm
	decided to do a very slight initial decrease with the hope that the congestion
	in transient it cannot know whether it was its decrease or PRR's decrease that
	was applied. It might be able to infer it from observations of the congestion
	window over CWR, but even if it is able to do that it has no way to change
	it.
\end{enumerate}

\subsubsection{PRR prevents rapid response}
In addition, PRR prevents DCTCP from reacting quickly to an increase in
congestion. DCTCP is designed to react to the extent of congestion and change
its reduction accordingly. However, when DCTCP deems that it needs to reduce
load quickly, PRR prevents it from doing so because it tries to do it
gradually. This can cause congestion to persist longer than necessary and
trigger new reductions that DCTCP without PRR would have been able to avoid.

\subsection{PRR Bug: Conclusion}
In this section we have shown a bug in the Linux's implementation of the
Proportional Rate Reduction (PRR) algorithm. We have explained why it happens
and the consequences of the bug. 
Specifically, the bug occurs in cases where i) the number of in-flight packets 
is less than the congestion window at the start of CWR (because of TSO deferral); 
and ii) there is a very small reduction of the congestion window. The bug leads
to a brief stall, followed by a burst once the congestion window has corrected itself.
This burst can trigger a new congestion event in the round following each reduction,
which in turn tends to cause under-utilization.

We have provided a simple change to the current code
that eliminates the bug and the associated issues. However, fixing the bug
uncovers an issue with fairness that we will look into in the rest of the report.

Whether or not PRR is appropriate for scalable congestion controls is
unclear. The idea of PRR is to gradually reduce the congestion window to the new
ssthresh over a round trip time. Scalable CCAs usually make very small adjustments to
their congestion window, and so the new ssthresh is typically very close to the previous
congestion window. In these cases when the reduction only needs to be small, there 
is little benefit in reducing gradually. And when the reduction needs to be relatively 
large it is questionable whether slowing it down is useful.

\clearpage
\section{Latecomer disadvantage}
\label{sec:late-comer}

\begin{figure}[b]
	\begin{minipage}{\linewidth}
\begin{lstlisting}[
		firstline=102,
		firstnumber=102,
		lastline=142,
		breaklines=True,
		style=cstyle,
		linebackgroundcolor={
			\ifnum \value{lstnumber} = 123 \color{pink}   \fi \ifnum \value{lstnumber} = 130 \color{orange}   \fi \ifnum \value{lstnumber} = 131 \color{orange}   \fi},
		caption={Source code of DCTCP in Linux, /net/ipv4/tcp\_dctcp.c},
		label={listing-dctcp-original}]
static u32 dctcp_ssthresh(struct sock *sk)
{
	struct dctcp *ca = inet_csk_ca(sk);
	struct tcp_sock *tp = tcp_sk(sk);

	ca->loss_cwnd = tp->snd_cwnd;
	return max(tp->snd_cwnd - ((tp->snd_cwnd * ca->dctcp_alpha) >> 11U), 2U);
}

static void dctcp_update_alpha(struct sock *sk, u32 flags)
{
	const struct tcp_sock *tp = tcp_sk(sk);
	struct dctcp *ca = inet_csk_ca(sk);

	/* Expired RTT */
	if (!before(tp->snd_una, ca->next_seq)) {
		u32 delivered_ce = tp->delivered_ce - ca->old_delivered_ce;
		u32 alpha = ca->dctcp_alpha;

		/* alpha = (1 - g) * alpha + g * F */

		alpha -= min_not_zero(alpha, alpha >> dctcp_shift_g);
		if (delivered_ce) {
			u32 delivered = tp->delivered - ca->old_delivered;

			/* If dctcp_shift_g == 1, a 32bit value would overflow
			 * after 8 M packets.
			 */
			delivered_ce <<= (10 - dctcp_shift_g);
			delivered_ce /= max(1U, delivered);

			alpha = min(alpha + delivered_ce, DCTCP_MAX_ALPHA);
		}
		/* dctcp_alpha can be read from dctcp_get_info() without
		 * synchro, so we ask compiler to not use dctcp_alpha
		 * as a temporary variable in prior operations.
		 */
		WRITE_ONCE(ca->dctcp_alpha, alpha);
		dctcp_reset(tp, ca);
	}
}
\end{lstlisting}
	\end{minipage}
\end{figure}

\begin{figure}
	\centering
	\includegraphics[width=.8\linewidth]{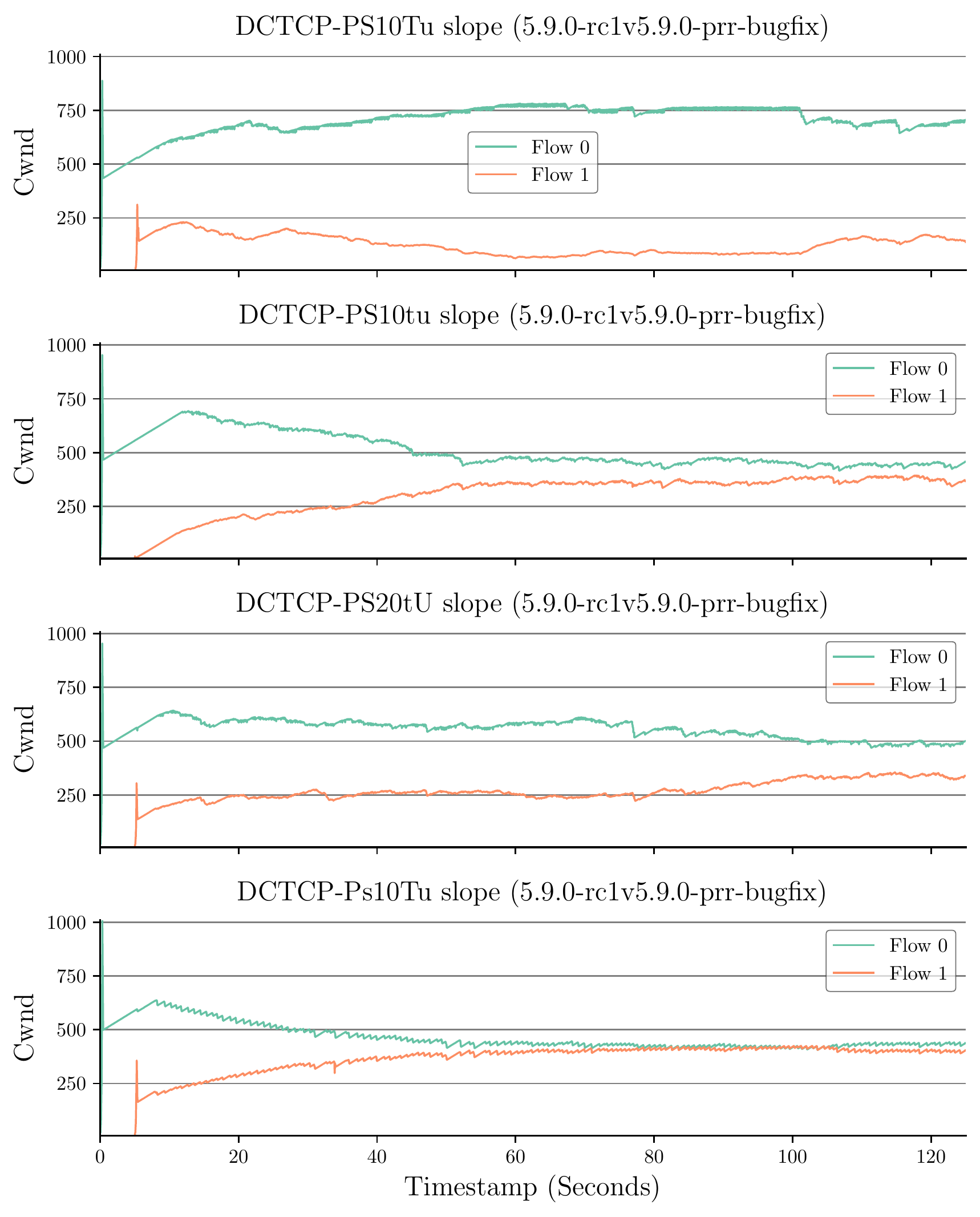}
	\caption{The latecomer disadvantage appears once the PRR bug is fixed.\\
		Time sequences
		of cwnd for two identical variants of DCTCP with starts staggered by 5\,s (all with PRR fixed): 
		i) regular DCTCP; ii) DCTCP without toggling to 0 of the EWMA; iii) DCTCP 
		with upscaled EWMA and 20-bit precision; iv) DCTCP without TSO. Each affects two flow's ability to converge to equal rate to different degrees.\\
		Capacity: 200\,Mb/s; RTT: 50\,ms; AQM: step at 2\,ms.}
	\label{fig:50ms200mbps-RFC-aligned-illustration-cwnd-two-flows}
\end{figure}

In certain scenarios, when the PRR bug presented in \cref{sec:prr-fix-align-rfc} is fixed, 
it unmasks a rate convergence problem between two identical long-running flows, as shown in
\cref{fig:50ms200mbps-RFC-aligned-illustration-cwnd-two-flows}. The top row is the current default implementation of DCTCP (but with the PRR bug fixed). The middle two rows show two different attempts to improve DCTCP's EWMA (explained below). Then the bottom row shows that the problem is related to TSO, because disabling TSO restores reasonable convergence. 

In the top three cases with TSO enabled (and the PRR bug fixed), it seems that the flows are falling into a local equilibrium before they converge. Initially, we conjecture that the bug in PRR
masked these convergence problems by introducing sufficient noise to jump the flows out of their local equilibrium into a more stable equilibrium with equal flow rates.

Before we try to untangle the cause of these results, we will introduce the two attempts to improve DCTCP's EWMA in the middle rows of \cref{fig:50ms200mbps-RFC-aligned-illustration-cwnd-two-flows}, with reference to the DCTCP EWMA code in \cref{listing-dctcp-original}. For those without in-depth knowledge of this code, a brief introduction is provided in \cref{appendix:dctcp-algo-and-implementation}. 
\begin{itemize}
	\item In the second row, `toggling' of the EWMA (line 123 of \Cref{listing-dctcp-original} coloured pink) is removed (T\(\rightarrow\)t). Without toggling, alpha floors at the relatively high value of 16 (the integer representation of 16/1024 or 1.6\%). With toggling, if alpha is less than 16 it is forced to zero.
	
	\item In the third row, as well as removing the toggling, alpha is upscaled (u\(\rightarrow\)U) and the EWMA precision is doubled from 10-bit to 20 (10\(\rightarrow\)20). The EWMA is upscaled by 16 so that low values can still be bit-shifted by 16 without being truncated to zero. This is the same approach as the EWMA of SRTT already in TCP. The combination of upscaling and higher precision ensures that low values of the EWMA down to about \(1/2^{20}\) or about 1E-6 can be represented in the variable alpha. Upscaling also fixes the problem in lines 30--31 of \Cref{listing-dctcp-original} (coloured orange), which otherwise round down the proportion of ECN feedback to the nearest 1/16, which black-holes ECN feedback completely if less than 1 in 16 packets in a round are marked.\footnote{\label{note:incr_deploy_DCTCP}Note that, once the toggle is removed from DCTCP, upscaling or improving precision would have to be done DC-wide, otherwise those flows still running with a high floor to the EWMA would starve themselves.}
\end{itemize}

\begin{figure}
	\centering
	\includegraphics[width=.8\linewidth]{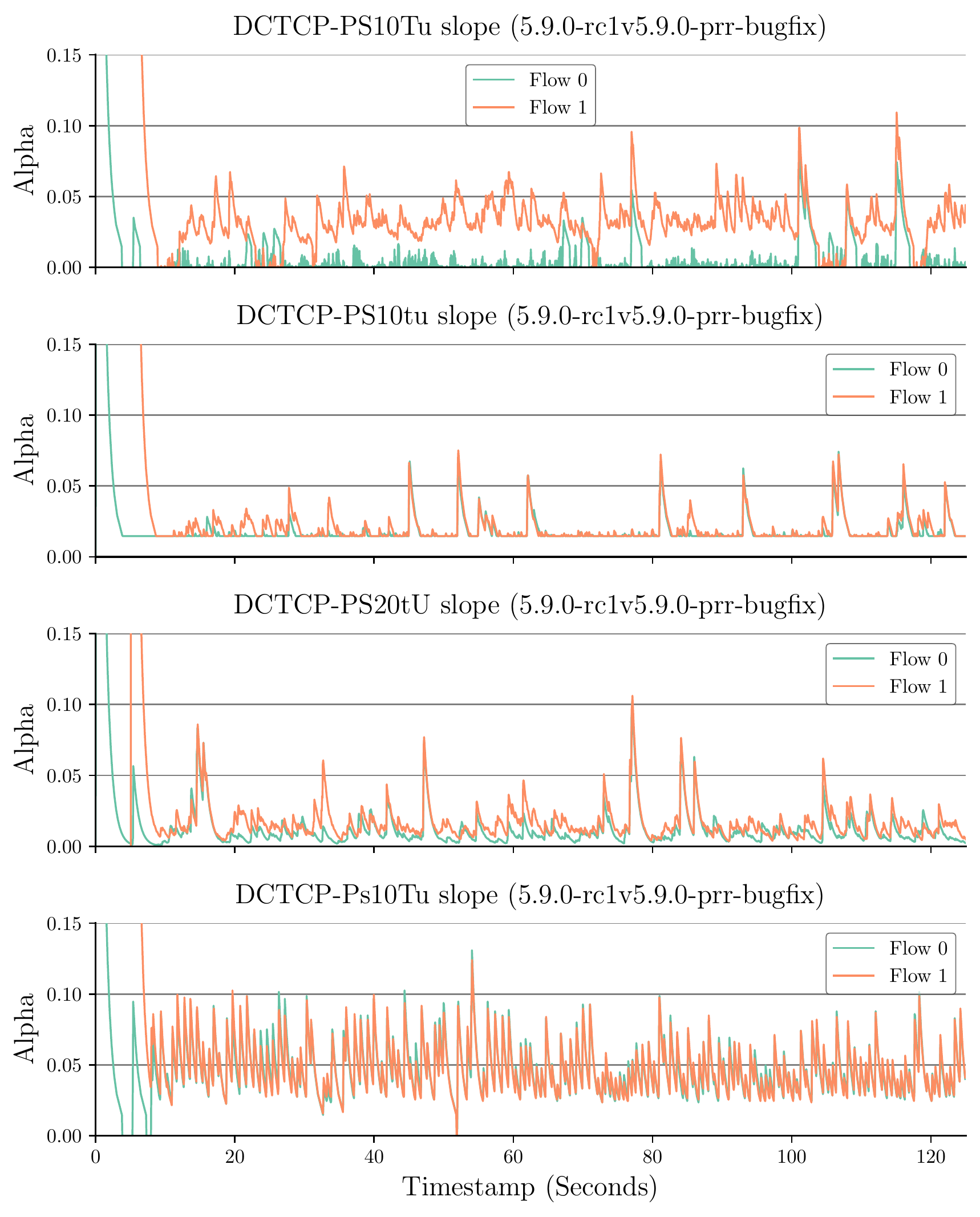}
	\caption{Time sequence of the DCTCP EWMA (alpha) for the two flows in \cref{fig:50ms200mbps-RFC-aligned-illustration-cwnd-two-flows}.} 
	\label{fig:50ms200mbps-RFC-aligned-illustration-alpha-two-flows}
\end{figure}

Returning to \cref{fig:50ms200mbps-RFC-aligned-illustration-cwnd-two-flows}, convergence does not seem to be such a problem
if toggling of alpha is removed (T\(\rightarrow\)t) and alpha
precision remains low (10) as in the second row. To explain this we can start by looking at the values of alpha for
the two flows in
\cref{fig:50ms200mbps-RFC-aligned-illustration-alpha-two-flows}.
We see that, when the flows converge to the same window, the alpha values
also converge to the same value. When TSO is disabled (last row, S\(\rightarrow\)s) the two flows quickly reach the
same alpha value, suggesting that they do get the same marking proportion from
their packets. This does not seem to be the case for the original default toggling behaviour along the top, nor for the higher precision upscaled DCTCP variant (third row). 

The `flooring' variant (second row) improves convergence in steady-state conditions, but we shall see later that it makes scalable CCAs lose their tight control under dynamic conditions.
The second row of \cref{fig:50ms200mbps-RFC-aligned-illustration-alpha-two-flows} clearly shows that alpha floors at 16, which is the
minimum representable value with a precision of 10 bits. So neither flow can reach the lower alpha value that would be expected at equilibrium. So, with toggling removed, each flow fools itself into thinking it is getting the same relatively high marking probability of 1.6\%, even if it is actually getting much less marking (which it should be at equilibrium). Therefore, its sawteeth vary much more than they need to. Nonetheless, flow rates converge reasonably well, but still not completely.

In the higher precision upscaled variant (third row), alpha now seems to converge to different values, which can only really occur if they get different marking
proportions from the network. It seems that the reason the flooring variant (second row) converges better is because flooring alpha mutes a difference
in marking proportion that something else is causing, which the higher precision variant (row 3) picks up on. However, forcing alpha to stick at a value higher than it is trying to reach doesn't really solve the problem, and it is likely to cause more queue variation than necessary. 

We dig more deeply into this in \cref{sec:late-comer-solutions}, after first checking behaviour with different RTTs.

\subsection{Dependency of latecomer disadvantage on RTT}
\label{sec:late-comer_RTT}

\begin{figure}
  \centering
  \includegraphics[width=.8\linewidth]{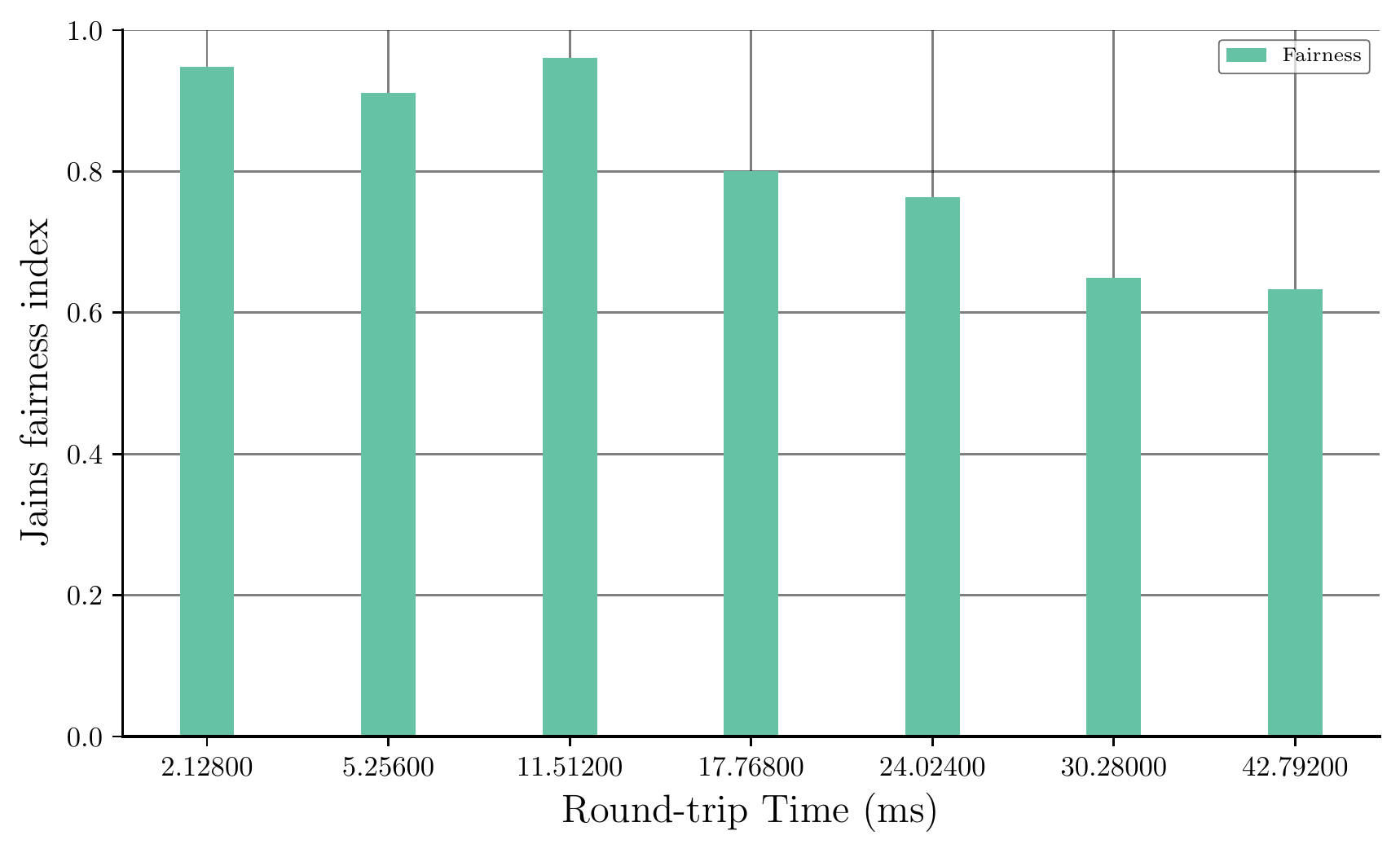}
  \caption{Latecomer disadvantage becomes more pronounced with RTT.\\
  	Rate fairness for
    two flows against their common RTT using two staggered-start identical regular DCTCP flows
    (DCTCP-PS10Tu). The latecomer always has the lower rate.\\
    Capacity: 200\,Mb/s; AQM: step at 2\,ms.}
  \label{fig:50ms200mbps-fixed-two-flows-illustration-fairness}
\end{figure}

\begin{figure}
  \centering
  \includegraphics[width=.8\linewidth]{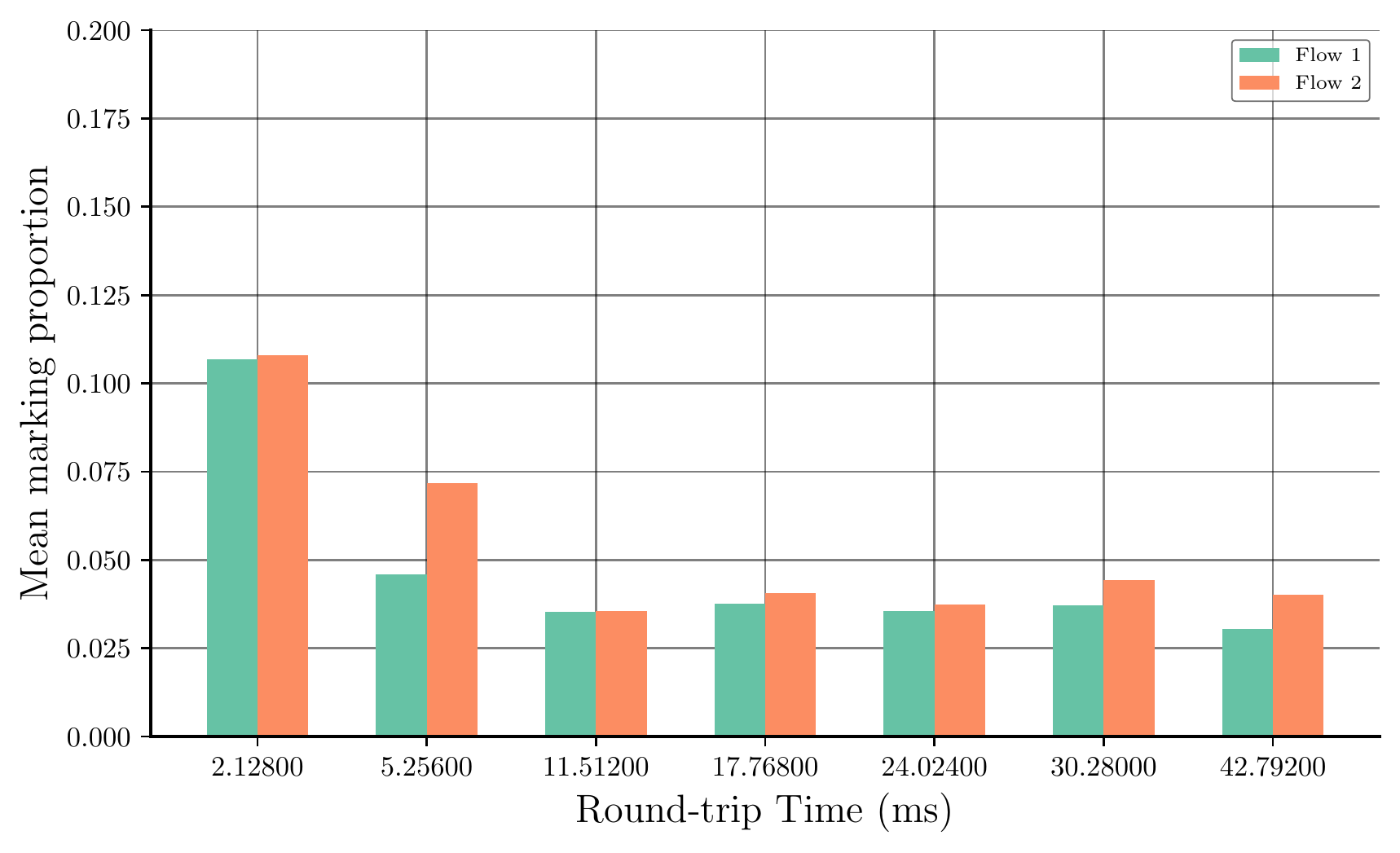}
  \caption{The latecomer (flow 2 with lower window) experience a higher
    marking proportion during the same experiment as \cref{fig:50ms200mbps-fixed-two-flows-illustration-fairness}.}
  \label{fig:50ms200mbps-fixed-two-flows-illustration-prop}
\end{figure}

At 200\,Mb/s the rate fairness between two identical flows varies with their common RTT as shown in
\cref{fig:50ms200mbps-fixed-two-flows-illustration-fairness}. Each flow's
measured marking proportion is plotted in
\cref{fig:50ms200mbps-fixed-two-flows-illustration-prop}. Notice that there is a
surprisingly small difference as the RTT increases, given the large
rate-unfairness. This is because many of the marks that flow 1 receives are not
used to reduce cwnd because alpha is toggled to 0 frequently. However, flows sharing the same queue would be expected to share the same marking probability. So, the
difference in marking probability experienced by the two flows suggests that the
cause of the latecomer disadvantage is not just in the CCA behaviour.

The Linux implementation of TSO tries to send one skb every 1\,ms. The number of
packets in the skb is based on the expected number of packets that the bottleneck can
process over 1\,ms. The expected number of packets is simply calculated as
\[
\frac{1 [ms] * W [pkt]}{SRTT [ms]},
\]
which is effectively the max burst in time multiplied by the packet rate.
Therefore, the flow with
the greatest packet rate will have the greatest burst size. That is the situation in
the experiments presented above. The later flow sends far smaller bursts while trying to push in than
the earlier established flow, because their rates are widely different. 
It is possible that the greater bursts are being marked proportionately less, so the
latecomer experiences a higher marking proportion and converges to a smaller congestion
window than the earlier, established flow.

We do not believe that this is solely an issue with the sender implementation. In \Cref{sec:altering-AQM} we show that there is a
more fundamental problem with how AQMs handle bursty traffic if their marking is based on sojourn time. But first we consider how much it is possible to mitigate the latecomer disadvantage at the sender alone. 

\subsection{Latecomer disadvantage: Potential remedies}
\label{sec:late-comer-solutions}

Five potential remedies for the latecomer disadvantage are introduced below. The last three are investigated further in the remainder of the report:
\begin{itemize}
	\item Disable TSO;
	\item Implement pacing in the NIC;
	\item Remove the toggling to zero of the EWMA;
	\item More tightly limit burst size;
	\item Alter the AQM.
\end{itemize}

\subsubsection{Disable TSO}
\label{sec:late-comer-solutions-disable-tso}
Disabling TSO removes the latecomer disadvantage, but is not a viable option
because the sender becomes processor-bound at multi-gigabit rates, resulting in
lower throughput.
DCTCP achieves roughly 3\,Gb/s over a 5\,Gb/s testbed bottleneck when TSO is disabled.
\job{Was this with the htb still configured for zero burst and shallow threshold?}
That is sending roughly one
1500B packet every 4 microsecond. That is processor time that is possibly taken
from other applications.
It is not just in data centers that this is an issue. Residential access capacity is
likely to continue to increase but packet sizes are much harder to increase. So, at some point TSO will probably become
necessary.

\subsubsection{Implement pacing in the NIC}
\label{sec:late-comer-solutions-pacing-in-NIC}
Pacing TSO bursts in the NIC would probably reduce or even eliminate the
latecomer disadvantage, but to our knowledge, few off-the-shelf NICs support 
packet pacing (yet).

\subsubsection{Remove the toggling to zero of the EWMA}
\label{sec:late-comer-solutions-remove}
The latecomer disadvantage doesn't show up in the DCTCP-PS10tu case with a floor to the EWMA in
\cref{fig:50ms200mbps-RFC-aligned-illustration-cwnd-two-flows} (second row). It can be seen in
\cref{fig:50ms200mbps-RFC-aligned-illustration-alpha-two-flows} that the floor to the EWMA prevents it reducing towards 0, which makes both flows fool themselves into thinking congestion is higher than it is. So whenever there is any congestion, they both react more than they need to and converge in large sawtooth steps, more like a Classic CCA. 

Although flooring the EWMA at 16 (instead of toggling to zero) largely removes the latecomer disadvantage, it causes 
other problems. Specifically, it loses the tight control loop that is meant to be a 
feature of scalable CCAs.
Also, it causes the limit cycle to grow as flow rate scales over the years. This means that the queue length measured in packets or bytes has to
increase with increased capacity to avoid underutilization. In contrast, a scalable
CCA should be able to preserve full utilization with the same size 
buffer (measured in bytes), so queueing delay should decrease as flow 
rate scales.

\begin{figure}
	\centering
	\includegraphics[width=.8\linewidth]{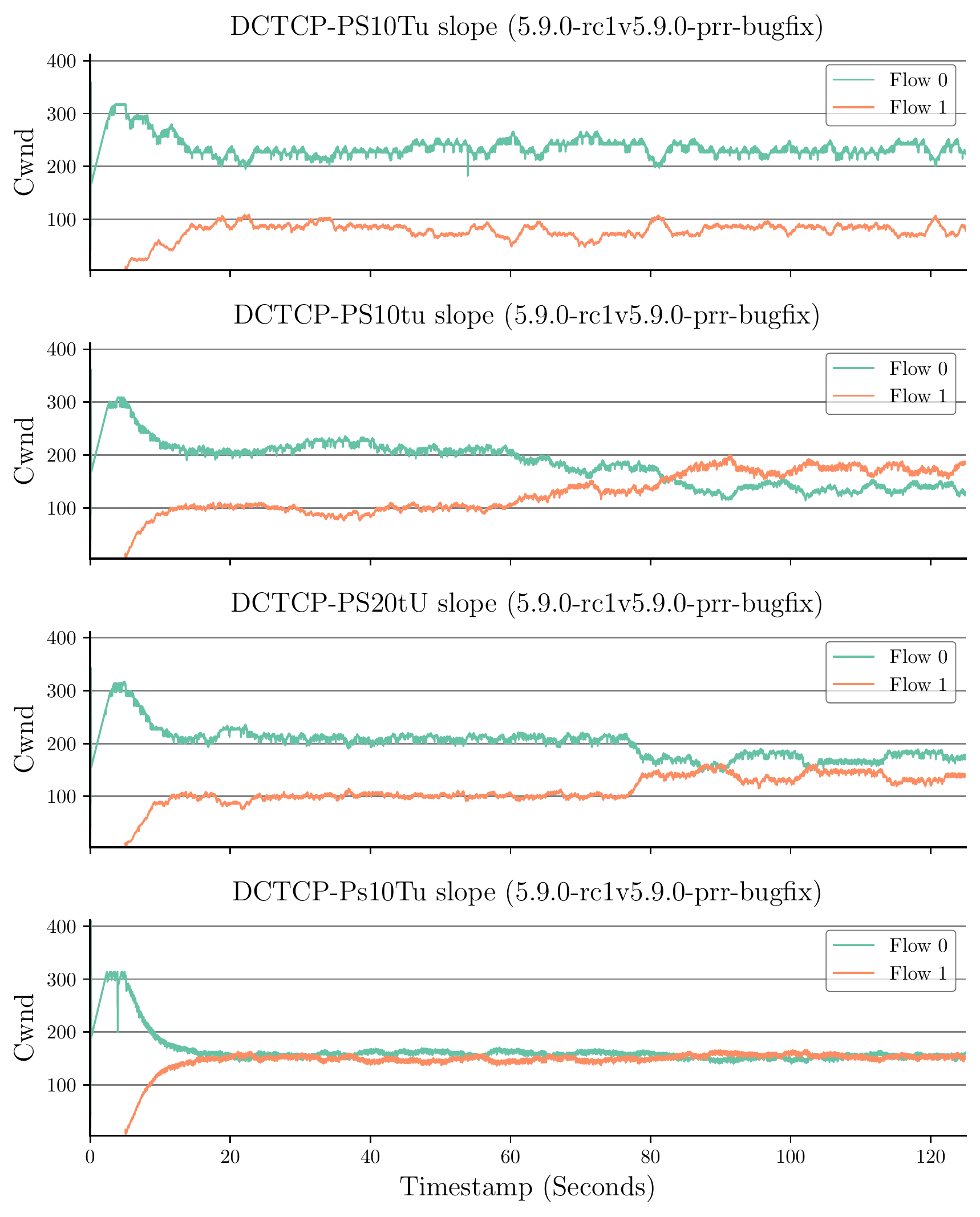}
	\caption{Removing the toggle of alpha is less effective at remedying slow convergence when RTT is smaller.\\
		\job{I would say convergence in the second row is no different.}
		Time sequences
		of cwnd with a smaller RTT but the same four variants of DCTCP as \Cref{fig:50ms200mbps-RFC-aligned-illustration-cwnd-two-flows} (all with PRR fixed): 
		i) regular DCTCP; ii) DCTCP without toggling to 0 of alpha; iii) DCTCP 
		with upscaled EWMA and 20-bit precision; iv) DCTCP without TSO.\\
		Capacity: 200\,Mb/s; RTT: 17.768\,ms; AQM: step at 2\,ms.}
	\label{fig:17ms200mbps-RFC-aligned-illustration-cwnd-two-flows}
\end{figure}

Also, removing toggling is not enough in scenarios where steady state
alpha is above the alpha's minimum value, for instance when BDP is low. One example using a low RTT is shown in
\cref{fig:17ms200mbps-RFC-aligned-illustration-cwnd-two-flows}. Convergence
without the toggle (the second row) is very slow.\job{It isn't any slower than with higher RTT.}

\subsubsection{More tightly limit burst size}
\label{sec:discussion-tcp_prague}
The unfairness issue is, at least partly, a result of difference in marking proportion
between the two competing flows caused by very different burst sizes. In the
DCTCP implementation the difference is amplified by toggling of alpha to 0;
It makes it harder for the flow with the bigger window to get away from alpha=0.

In the scenarios we have tested (see \cref{sec:altering-AQM}), TCP Prague largely avoids the latecomer disadvantage.
In TCP Prague, alpha has 20 bit precision and it is stored as an upscaled value. In
addition it does not toggle alpha to 0. Therefore Prague should have similar behaviour to DCTCP-PS20tU. However, we see later (\Cref{sec:toggling-issue}) that DCTCP-PS20tU does less well at avoiding the latecomer disadvantage. 

The main other difference is that DCTCP uses Linux's default max burst duration of 1\,ms, whereas 
TCP Prague limits bursts to 250\,\(\mu\)s. So limiting burst size to significantly less than the marking threshold seems an important factor.%
\footnote{Note that, as the bottleneck rate increases and approaches the rate of the sender's NIC, the size of the max burst that will cause 250\,\(\mu\)s of queuing becomes an under-estimate if it is calculated from 250\,\(\mu\)s * cwnd / RTT. This has to be borne in mind when interpreting results --- if no latecomer disadvantage is apparent when measuring data centre scenarios, it could simply be because bottleneck link rates in DCs are closer to NIC rates. So the max burst size that DCTCP calculates to avoid 1\,ms of queuing could actually cause much less than 1\,ms of queuing.} 
This is investigated further in \cref{sec:altering-AQM}.


\subsubsection{Alter the AQM}
\label{sec:late-comer-solutions-changing-aqm}

In recent work on rapid signalling of queue dynamics~\cite{Briscoe17b:sigqdyn_TR}, it was noticed that using the sojourn time of the packet being dequeued to measure queue delay exhibits two problems: i) it fails to use the latest queue information, instead measuring the delay of the queue as it was when the packet arrived; ii) it marks a packet based on the delay it experiences itself, rather than the delay it causes to other packets backlogged behind it. 

On further investigation, this latter point was found to be particularly problematic with traffic bursts that cause delay of the same order as the marking threshold. Then, most or all the packets within the burst would not be marked, but any packets arriving more smoothly would be likely to sit behind the burst as it drained, and therefore more likely to be marked themselves.

Therefore, in \cref{sec:altering-AQM} we shift attention from the congestion control at the sender to investigating a possible alteration to the AQM in the network.

\clearpage
\section{Removing the Toggle of Alpha to Zero}
\label{sec:toggling-issue}

In this section we will discuss why the toggling of alpha to 0 on line 123 in
the current (Sep 2021) DCTCP Linux implementation (\Cref{listing-dctcp-original} in \Cref{appendix:dctcp-algo-and-implementation}) is believed to be problematic.\footnote{This issue/bug was introduced in Nov 2015~\cite{shewmaker15:Linux_DCTCP_EWMA}, being first applied in the v4.3 kernel. The bug was originally discovered by Koen De Schepper and a patch submitted, but not accepted on the grounds that no problem had been apparent. This motivated the start of work on the present report (before the full extent of the tangle of problems was understood).}

Line 123 executes the first part of the EWMA, where alpha
is subject to a decay. The function min\_not\_zero returns the minimum of two
values, but if either is equal to zero the other one is returned. Zero is
returned if both are zero. Consequently, if the decay should be smaller than
that allowed by the granularity of alpha, the second term will become 0 and the
current value of alpha is returned by min\_not\_zero. This causes alpha to be
reduced to 0, effectively fast-forwarding the decay calculation. That means that alpha
values between $1/1024$ and $15/1024$ are unused. We call this range a deadzone.

The use of min\_not\_zero was accepted on very weak grounds in commit
c80dbe0461298. It states that it is a problem that dctcp\_alpha can never reach
0. It claims that this 'could gradually drive uncongested flows with leftover
alpha down to cwnd=137'. No empirical evidence seems to have been required to support this claim. In practice, this would only be possible if there were congestion marks in most round-trips (perhaps due to bursty flows caused by TSO or smaller flows). Otherwise there would be nothing to trigger reductions in cwnd, no matter the value of alpha. Even if there were some marks caused by burstiness, the more cwnd reduced, the lower the likelihood of any further marks in each round. 

\subsection{Toggling alpha to zero: Problem}
\label{sec:toggling-problem}

The toggle to zero effectively makes a flow unresponsive until
congestion gets bad enough to push dctcp\_alpha above 15. If there is a
continuous
arrival of smaller flows one or more existing flows might not yield spare
capacity to them. This causes the load of the smaller flows to be absorbed by
the bottleneck's queue, introducing queueing delay spikes.

This deadzone from 0 to 15 is problematic because it means that DCTCP is less scalable than it
should be. As we discussed in \cref{sec:scalablity}, DCTCP's scalability property
holds if alpha stays at its equilibrium value. As capacity grows alpha has to
decrease to keep DCTCP scalable. However, the current implementation cannot
achieve equilibrium alpha values for congestion windows larger than 128.\footnote{Given the equilibrium fraction of CE marks reduces as flow rate scales, if the two marks expected at equilibrium are fed back, the algorithm will always black-hole them both for flow rates with \(\mathrm{cwnd} > 128\). And it will remain unresponsive until it forces the number of CE marks to rise to the next integer above \(\mathrm{cwnd}/64\) (when it will register just 1 mark per RTT) and black hole the rest.}
The toggling of dctcp\_alpha is not the cause of this. The original code that floored at 16 was the cause, but toggling to zero is not the right
remedy. 

Removing the toggling to zero is not necessarily a sufficient remedy either. Even if the line with the min\_not\_zero function is removed, when the current DCTCP code feeds the count of CE markings into the EWMA at the end of each round, it rounds down the fraction of markings to an integer multiple of 1/64 (including zero). This creates another dead zone between zero and 1/64 (1.6\%). This time, the dead zone is not during EWMA decay, but during input of congestion markings. 

This flaw is in lines 130--131, still referring to the dctcp\_update\_alpha function in \cref{listing-dctcp-original}, where they are coloured orange. As already explained, in integer arithmetic, the current DCTCP represents dctcp\_alpha by integers in the range 0--1024. At line 130, delivered\_ce is therefore multiplied by 1024, but at the same time it is factored down by the gain of the EWMA (\(1/16\)), which is achieved by the left-shift by \(-\)dctcp\_shift\_g, which is -4. 

For instance, if flow rate is 200\,Mb/s and RTT=40\,ms; cwnd = 667 so on average delivered = 667, then if there are 8 CE marks in a round, delivered\_ce = 8, and the two lines result in:\\
\(130: delivered\_ce = 8 << (10-4) = 512;\)\\
\(131: delivered\_ce = 512/667 = 0;\)\\
So 8 CE marks get black-holed. Upscaling the EWMA is intended to address this flaw. It is investigated below, but with unexpected results.

\subsection{Toggling alpha to zero: Evaluation}
\label{sec:toggle_eval}

In the following experiments we highlight the difference in cwnd and
dctcp\_alpha evolution for implementations that handle the update of
dctcp\_alpha differently. We run the experiments first with a step threshold at 2\,ms
then with a linear ramp from 2\,ms to 4\,ms.

%
%
For these experiments the capacity is set to 200\,Mb/s, and the RTT is 91.8\,ms.
The values were chosen to create a large
congestion window which shows the dynamics more clearly. In addition, this allows
the improved dynamics to feed through as improved utilization.

\begin{figure}
  \centering
  \includegraphics[width=.8\linewidth]{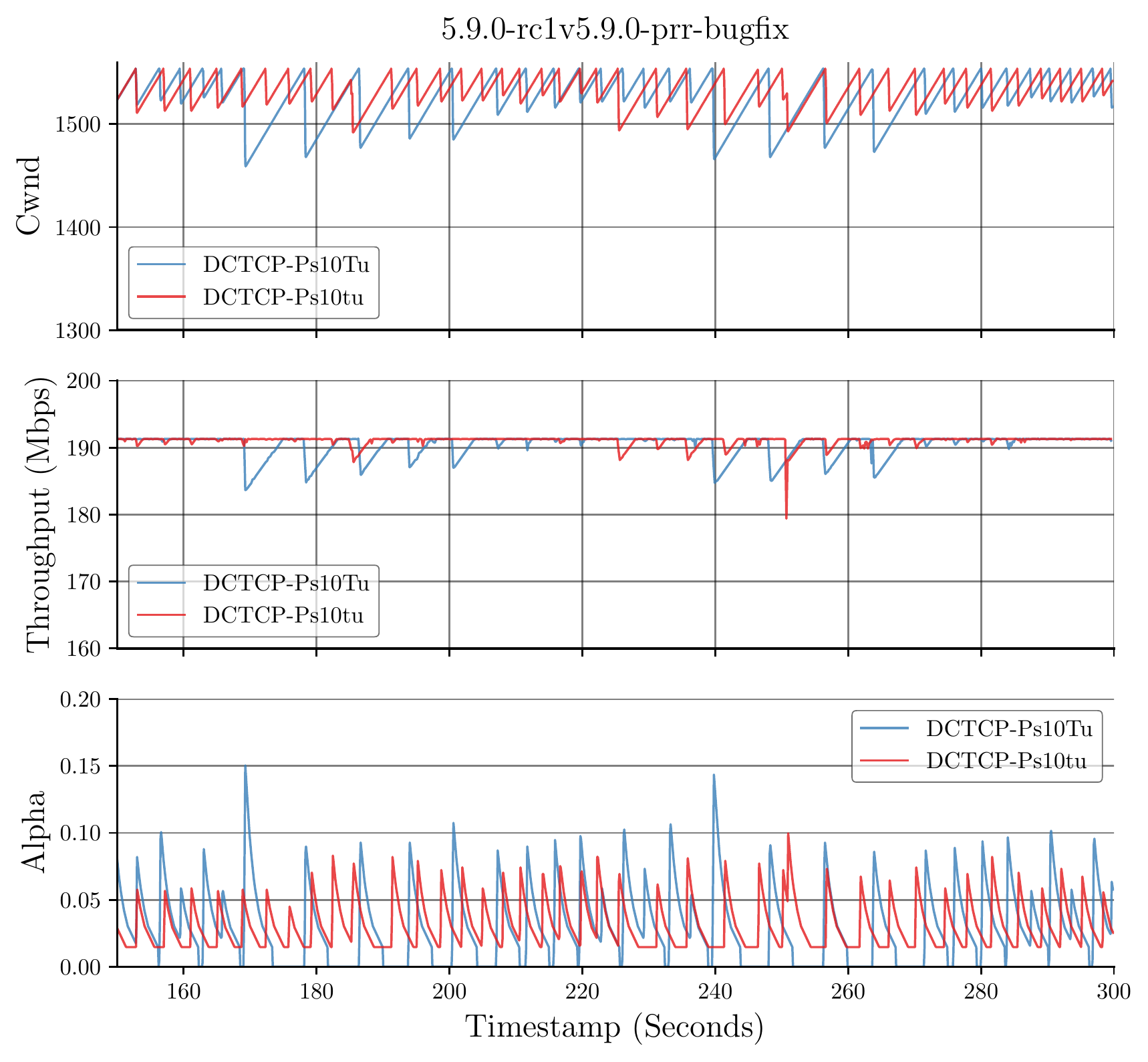}
  \caption{Toggling of alpha to 0 (blue) increases the chance of underutilization 
  	due to back-to-back reductions.\\
    The plot shows cwnd, throughput and alpha time-series for
    two different DCTCP
    implementations: with and without toggling of alpha to 0 (T \& t).\\
    Capacity: 200\,Mb/s; RTT: 91.8\,ms; AQM: step at 2\,ms.}
  \label{fig:toggle:step_toggle_vs_non-toggle}
\end{figure}

\subsubsection{Step at 2\,ms}
\label{sec:toggle_eval_step}

\Cref{fig:toggle:step_toggle_vs_non-toggle} plots the evolution of cwnd,
throughput and alpha for DCTCP with and without toggling of dctcp\_alpha to 0
when it reaches 15.

With toggling, the flow seems to experience
regular consecutive congestion events, one when alpha is 0 and one when alpha
has been increased enough for a reduction to occur. The flow without the
toggling seems to relieve congestion through a single reduction and has less
oscillation in alpha and cwnd. Throughput also remains higher when dctcp\_alpha
is not toggled to 0.

\begin{figure}
  \centering
  \includegraphics[width=.8\linewidth]{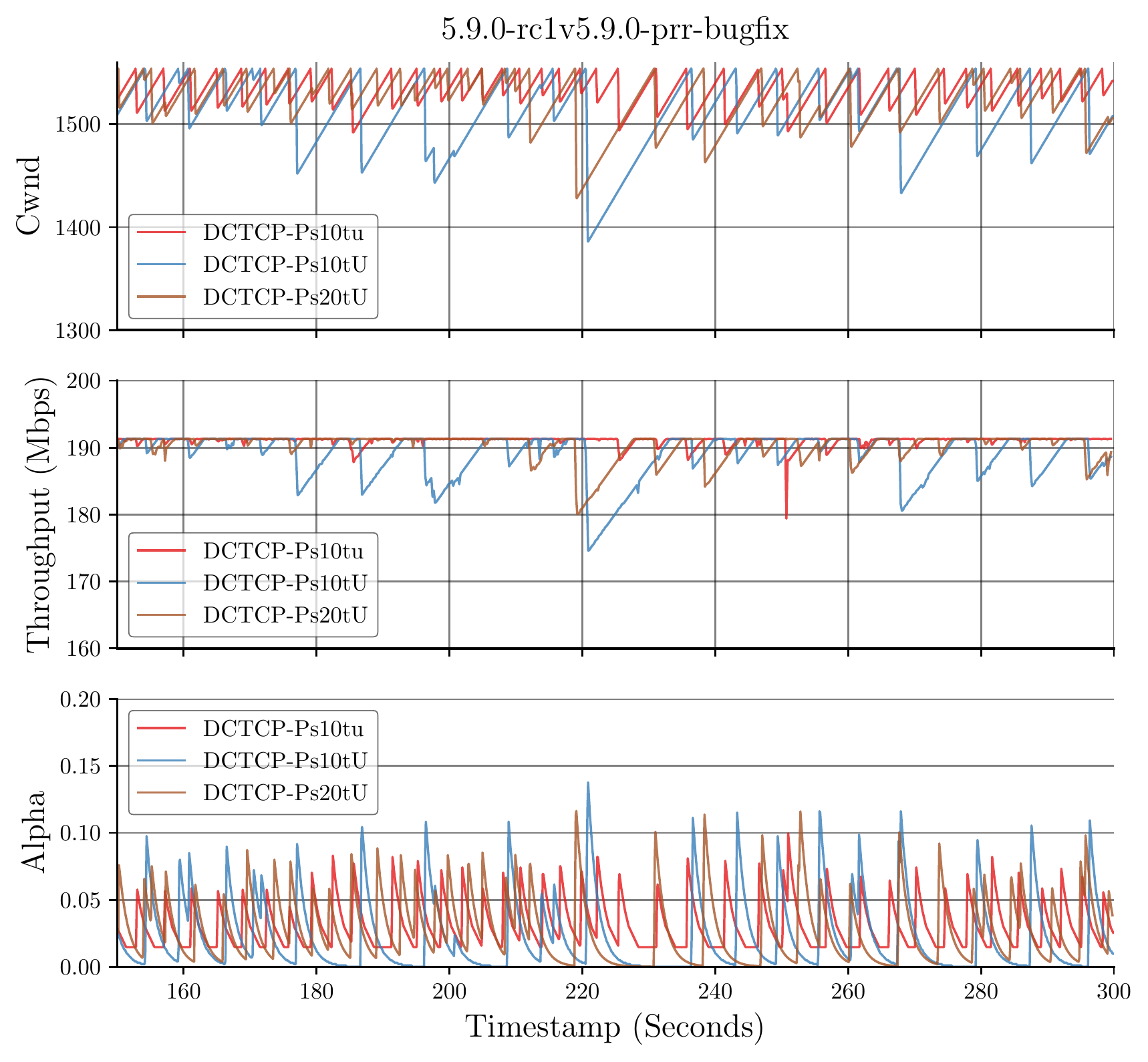}
  \caption{Once toggling to zero has been removed (t), upscaling (U) seems to make utilization worse, perhaps by increasing the chance of back-to-back reductions
  	\job{This previously said, ``higher precision reduces the chance of back-to-back reductions and
    	underutilization, but it is still sensitive''} 
    A high floor to the EWMA of \(16/1024\) (red) seems to make it less sensitive to noise,
    by effectively becoming a classic CCA with a fixed lower bound to its multiplicative decrease. 
    The plots show cwnd, throughput and alpha time-series for three different DCTCP
    implementations: without toggling of alpha to 0 (DCTCP-Ps10tu), alpha stored
    upscaled with 10-bit precision (DCTCP-Ps10tU), and alpha stored
    upscaled with 20-bit precision (DCTCP-Ps20tU). The red plot is the same as the red plot in \cref{fig:toggle:step_toggle_vs_non-toggle} for comparison.\\
    Capacity: 200\,Mb/s; RTT: 91.8\,ms; AQM: step at 2\,ms.}
  \label{fig:toggle:step_upscaled_vs_non-toggle}
\end{figure}

\Cref{fig:toggle:step_upscaled_vs_non-toggle} plots the evolution of cwnd,
throughput and alpha for three variants of DCTCP: i) without toggling of alpha; ii) alpha stored upscaled
without changing the 10-bit precision; and iii) alpha stored upscaled with 20-bit precision.
The upscaled versions make the values below 16 available to dctcp\_alpha, and
should therefore allow for more gradual adjustments of cwnd and less
oscillation.

However, that seems to not be the case. We conjecture that this is due to the use of a step AQM.
A step
makes the marking probability switch between 0 and 100\% marking for at least a full round until the response can reduce the queue back below the threshold. This causes a large window reduction, so additive increase then takes many rounds to recover the lost ground. The result is much more like a Classic large-sawtooth variation in the window and consequently leads to high variation in queuing delay.

In the next experiments, we try switching
from a step to a ramp, expecting to make the low values of alpha more accessible. It might then even be possible to increase the smoothing factor (gain) to
make response to a change in congestion faster.\footnote{However, investigation of that is set aside for further work, because it would also reduce a long flow's memory of congestion due to background short flows.}

\clearpage
\subsubsection{Linear ramp from 2\,ms to 4\,ms}
\label{sec:toggle_eval_ramp}

\begin{figure}
  \centering
  \includegraphics[width=.8\linewidth]{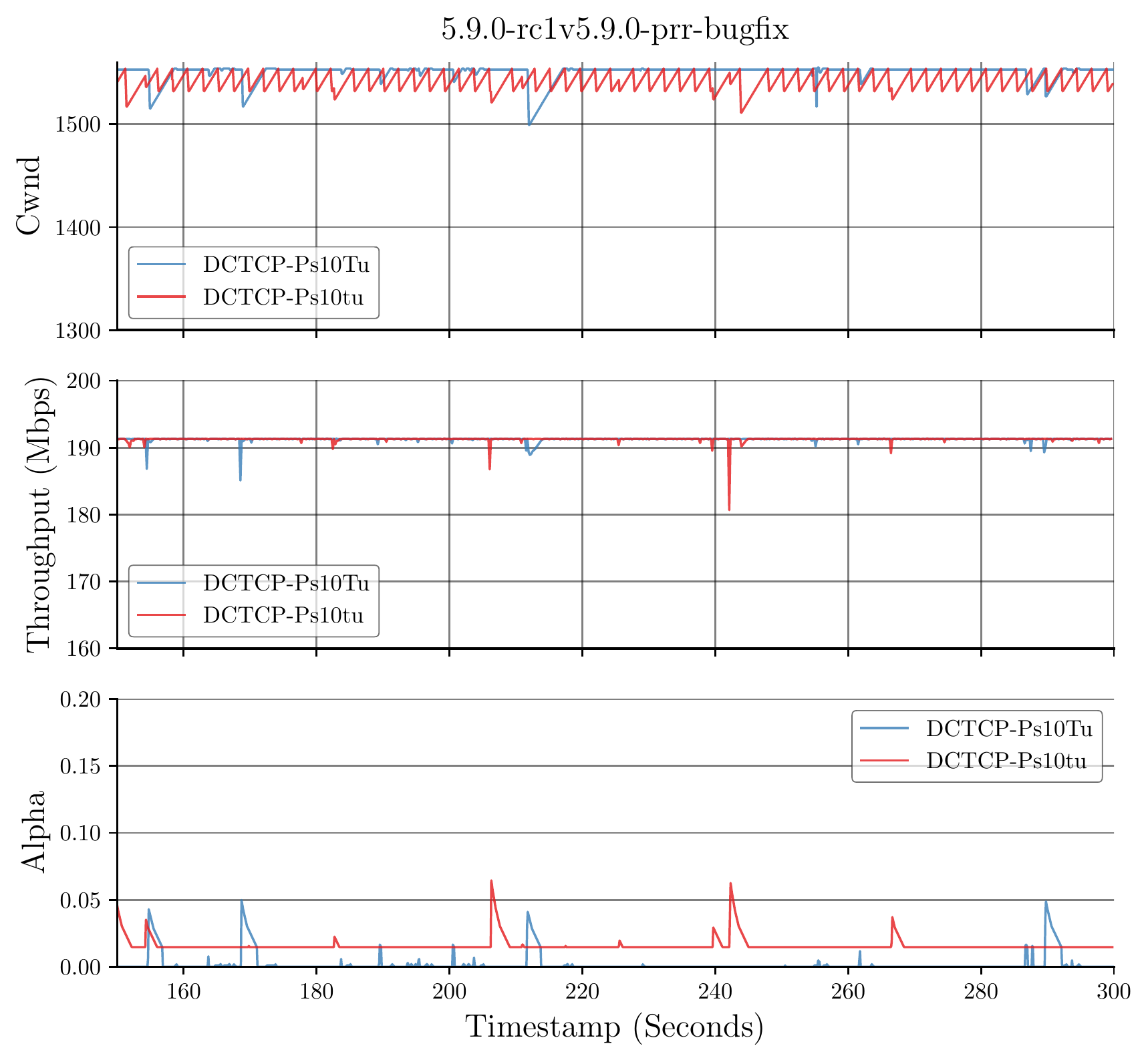}
  \caption{With a ramp AQM and a DCTCP implementation that toggles alpha to 0 (blue) the initial reaction to a congestion signal
    is to not reduce cwnd at all. In this steady-state experiment the user will not
    experience degraded performance (utilization is 100\%), but the text explains that flooring alpha will lose the potential advantages of a scalable congestion control in dynamic scenarios.\\
    The plots show cwnd, throughput and alpha time-series for two different DCTCP
    implementations: with and without toggling of alpha to 0 (T \& t).\\
    Capacity: 200\,Mb/s; RTT: 91.8\,ms; AQM: ramp from 2\,ms to 4\,ms.}
  \label{fig:toggle:ramp_toggle_vs_non-toggle}
  \job{Throughout, it's not clear whether there's 100\% utilization, because the throughput quoted in the charts is presumably without some headers, so it doesn't match the Capacity quoted. Joakim will check whether it's easy to convert to utilization by accounting for header sizes.}
\end{figure}

\begin{figure}
  \centering
  \includegraphics[width=.8\linewidth]{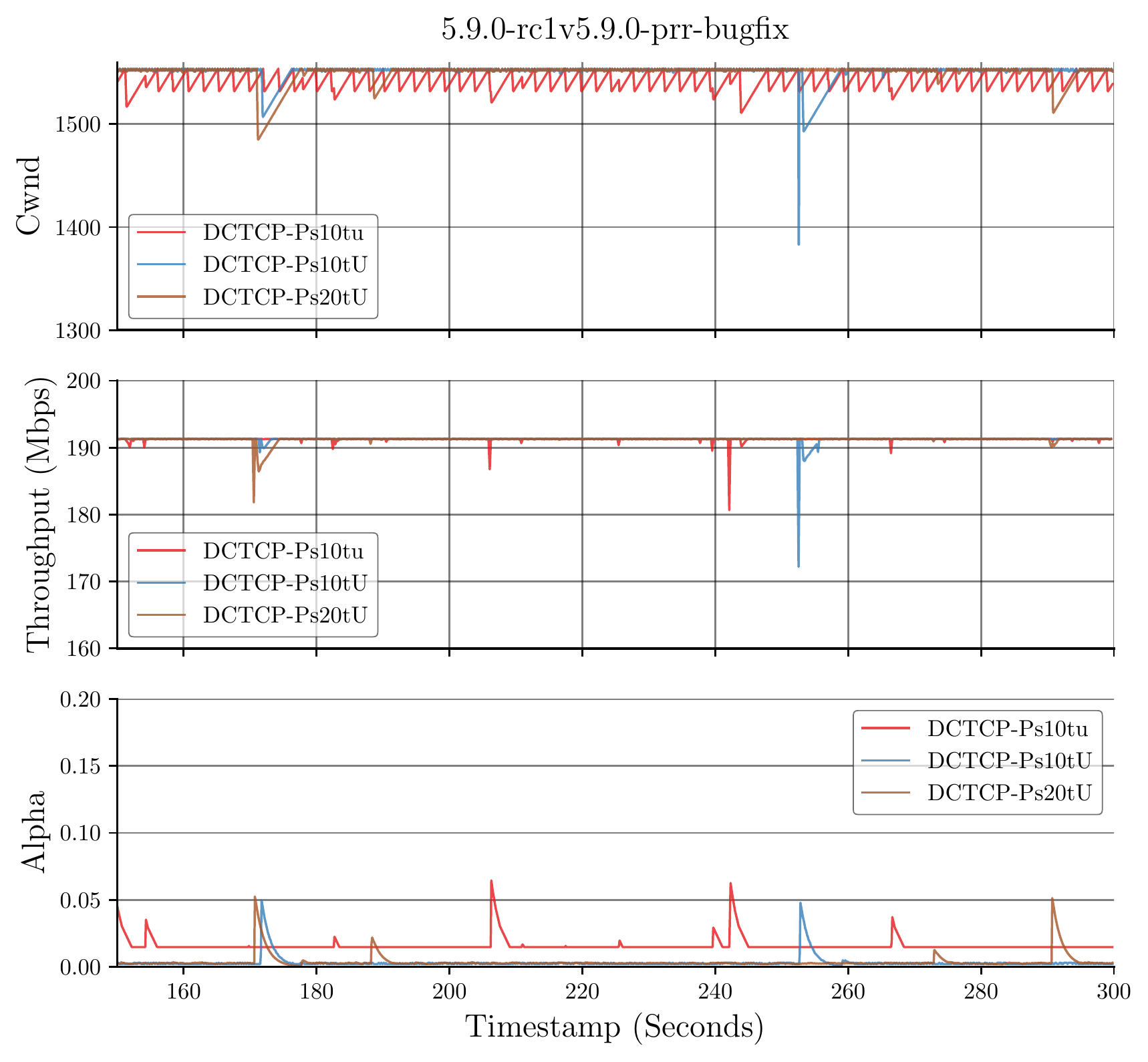}
  \caption{With a ramp AQM, an upscaled EWMA (blue or brown) reduces steady state oscillation, but makes
    it more susceptible to noise, which can cause occasional back-to-back reductions.
    The plots show cwnd, throughput and alpha time-series for same three DCTCP
    implementations: as in \cref{fig:toggle:step_upscaled_vs_non-toggle}, but with a ramp AQM. The red plot is the same as the red plot in \cref{fig:toggle:ramp_toggle_vs_non-toggle} for comparison.\\
    Capacity: 200\,Mb/s; RTT: 91.8\,ms; AQM: ramp from 2\,ms to 4\,ms.}
  \label{fig:toggle:ramp_upscaled_vs_non-toggle}
\end{figure}

An AQM
that can provide more fine grained feedback is a solution to the large Classic-like sawteeth with a step. The simplest possibility is a linear ramp that marks packets proportionally to how far up the
ramp the current queueing delay is (rather like RED, but configured using time units, and with no queue averaging). \cref{fig:toggle:ramp_toggle_vs_non-toggle} \& \cref{fig:toggle:ramp_upscaled_vs_non-toggle} show that a ramp enables smaller
marking proportions so that the window oscillations can become completely smooth to the eye, except in the two cases where the EWMA in the sender prevents itself from representing smaller marking proportions, as shown in the two red cases with a low precision floor to the EWMA (.\,.10tu). 

Step marking also induces synchronization between the window sawteeth of flows. Without synchronization, the amplitude of the queue delay variation decrease as the number of flows increases. But, with synchronization, the size of the combined oscillation tends to the sum of all the synchronized individual variations.

In the following experiments the AQM uses ramp marking where the probability grows linearly from 0 to 1 as queuing delay grows from 2\,ms to 4\,ms.

\Cref{fig:toggle:ramp_toggle_vs_non-toggle} plots the evolution of cwnd,
throughput and alpha for DCTCP with and without toggling of dctcp\_alpha to 0
when it reduces below 16.

Here we see an interesting situation where the congestion window stays almost constant
when dctcp\_alpha is toggled to 0.
This is because 
the flow with alpha at 0 does not reduce its window,
and Linux DCTCP does not increase its window either, because it does not do additive
increase while in CWR state for a round after a reduction, and it resets its cumulative count (snd\_cwnd\_cnt) upon
entering CWR. 
In effect, the flow is stuck. This seems great, as
throughput stays high and there is no oscillation, but in a competitive scenario
this prevents convergence. Two or more flows that suppress additive increase during CWR and toggle alpha to zero can all get stuck in a
local equilibrium when congestion is less than that needed for alpha to exceed 15, for instance where each flow gets a couple of marks each round-trip time. Then they all suppress
both additive increase and multiplicative decrease.%
\footnote{The Prague congestion control continues additive increase throughout CWR state, so it doesn't experience such a complete lack of sawtoothing. It also helps that Prague carries over any remaining snd\_cwnd\_cnt properly. Specifically, Prague divides its increase over the ACKs arriving in a round, but suppresses the increase on ACKs that report congestion feedback, which ensures that additive increase still reduces as congestion increases. Ironically, continuing AI during CWR was originally added to Prague to make window oscillations smoother, because suppressing AI during CWR left little time for any increases at all unless reductions were large.}

When dctcp\_alpha is not toggled to 0 it remains floored at 16 which causes the same
predictable oscillation as when a step threshold is used, but it seems to 
mitigate the latecomer disadvantage problem. However, as flow rate scales, this 
effectively fixes DCTCP's window reduction whatever the extent of congestion 
(unless congestion is really high).%
\footnote{Indeed, flooring alpha at 16 effectively turns DCTCP into a hybrid between DCTCP and ABE (Alternative Backoff with ECN~\cite{Khademi18:ABE}) with a very low but fixed decrease factor of 1/16.}
So DCTCP would no 
longer be able to respond to the extent of congestion marking effectively, 
and it would no longer induce frequent marking. So it would lose the tight 
control of a scalable CCA. For instance, it would no longer be able to rapidly detect when congestion marking had stopped.%
\footnote{Flooring alpha at 16 is essentially a hack to mitigate a problem with the way DCTCP and Prague trigger a reduction on the first CE mark (when alpha has often decayed to zero), then suppress any further reduction for a round of CWR state, just when most congestion feedback is arriving. A technique to measure and respond to congestion per-ACK rather than per-RTT is in the process of evaluation~\cite{Briscoe22a:Prague_clock_lag}. it also removes all the unnecessary lag in the EWMA clocking machinery of Prague, which adds up to between 1 and 2 round trips. However, although that is part of the tangle of problems with the Linux implementation of DCTCP, it involves such a significant design change that we have had to draw a line and place it beyond the scope of the present report.}

\Cref{fig:toggle:ramp_upscaled_vs_non-toggle} plots the evolution of cwnd,
throughput and alpha for DCTCP without toggling of alpha, alpha stored upscaled
with 10 bit precision and alpha stored upscaled with 20 bit precision. In all cases the AQM is a ramp from 2\,ms to 4\,ms.

Here, we see the benefit of storing alpha with increased precision as an upscaled
value. Alpha is able to stay at and reach a small area around its equilibrium
value. This causes very little oscillation in general, but we also see that a
small amount of noise can make it get consecutive reductions. Such reductions
happen when there is no reduction or the reduction is too small to alleviate
congestion. The magnitude of the reductions suggest that there is some noise
causing a rapid increase in congestion. The smaller oscillation in alpha and
cwnd makes it more sensitive to noise because it picks up on it more
easily. When there is a regular oscillation, as when toggling is off, noise can
hit at different times in the oscillation cycle. If it hits when it is at the
bottom, the flow might not notice it. However, with increased precision, a flow
will detect and react to it.

The DCTCP paper justifies using a step on the basis that, 
\begin{quote}
	`Prior work \cite{Raina05:CtrlTheory_BufferSizingPtII}, \cite{Kelly08:StabilityCCSmallBuffers} on congestion control in the small buffer
	regime has observed that at high line rates, queue size fluctuations
	become so fast that you cannot control the queue size, only its distribution. The physical significance of \(\alpha\) is aligned with this observation: it represents a single point of the queue size distribution at
	the bottleneck link.'~\cite{Alizadeh2010:DCTCP_Short}
\end{quote}
However, just saying that a ramp might be no better than a step does not 
prove that a step is as good as a ramp, and in practice using
the on-off signal from a step makes the EWMA unnecessarily slow to converge on an 
average of an ever-changing congestion signal. It also takes much longer to detect that the signal has disappeared after available capacity increases, because the on-off cycle makes it normal for the signal to disappear for tens (or even hundreds) of rounds at a time (about 24 rounds in \cref{fig:toggle:ramp_toggle_vs_non-toggle} \& \cref{fig:toggle:ramp_upscaled_vs_non-toggle}).

If it is inevitable that a ramp will appear increasingly step-like as flow rate scales, it might be more fruitful to `emulate a ramp' at the sender. This can be done by dithering the smoothness of packet pacing to induce a more gradual increase in marking as the queue approaches the step threshold. This is beyond the scope of the present paper, but worth further investigation.

\subsection{Interim Conclusions}
\label{sec:toggling-solution}

A partial improvement to the latecomer disadvantage is to floor dctcp\_alpha 
at 16 (representing 16/1024 or 1.6\%) by removing the toggle to zero at line 123 in \cref{listing-dctcp-original}. To an extent, this improves the otherwise poor
convergence rate and ensures that DCTCP reduces its unfairness against latecomers, but at the expense of considerably more steady
state oscillation and possible under-utilization.
Also, as flow rate increases over the years, flooring the EWMA at 16 
will make the time between congestion marks continually increase, in turn making
DCTCP's congestion control less responsive to change (i.e.\ unscalable).

To reduce the window oscillations, it is possible to 
make the values below 16 reachable by storing dctcp\_alpha in an
upscaled variable, shifted by dctcp\_shift\_g (as implemented in the Prague CCA). Then alpha\_value can be scaled down whenever it is used. Even if dctcp\_alpha is able to use values from 1 to 15 it will still have an issue
as capacity scales, because alpha values below $1/1024$ will not be
not representable. This can be fixed by increasing the precision of alpha, 
which can be a parameter to the module that can be adjusted in
future if scalablity becomes an issue.

This upscaled and higher precision EWMA addressed the smoothing, but it was disappointing to find that it did not address the latecomer disadvantage as effectively as just flooring dctcp\_alpha at 16. It is believed this is a symptom of an interaction between the burst sizing in the Linux sender and burst marking in the step marking in Linux. This interaction focuses more marking on lower rate flows, which causes the latecomer to reach a local equilibrium before it has converged. This is investigated further in the next section (\cref{sec:altering-AQM}). 

We conjecture that flooring alpha at 16 appears to partially solve the latecomer disadvantage because it fools all flows into thinking they are experiencing the same relatively high marking probability, thus masking the marking bias against lower rate flows. Flooring alpha effectively turns DCTCP into a classical congestion control, which applies a relatively large, fixed cwnd reduction on the first sign of congestion. Although mimicking a Classical CCA avoids the latecomer disadvantage, it also loses all the rapid control advantages of scalable CCAs, which are not needed in the steady-state scenarios of the present paper, but they come to the fore in dynamic scenarios.

One thing seems absolutely clear, the toggling patch applied to DCTCP in 2015 has no merit and should be reverted.

It is also clear that congestion marking at a step threshold causes cwnd to vary
excessively --- much as a Classical CCA would. Using a ramp AQM instead results
in a very smooth cwnd trace. Therefore a ramp AQM seems preferable, or it might
be sufficient to use a step in the network but synthesize a gradual increase in
marking as the queue grows by dithering pacing at the sender.

Via footnotes, we have also recommended investigation into two other changes to DCTCP's EWMA processing, which are being assessed separately because they are beyond the scope of this report:
\begin{itemize}
	\item Continuing additive increase during CWR state (except on ACKs carrying congestion feedback), as implemented in the Prague CC;
	\item Per-ACK clocking of the EWMA and per-ACK reduction of cwnd during CWR state~\cite{Briscoe22a:Prague_clock_lag}.
\end{itemize}
Both these proposals illustrate how the change from an existence-based to extent-based response to congestion needed a more radical overhaul of the DCTCP implementation, which still needs more of the classical machinery stripped out.

\clearpage
\section{Interaction between Scalable CCA's \& AQM's}
\label{sec:altering-AQM}

The earlier experiments with various improvements on the DCTCP EWMA mitigated but could not eliminate the latecomer disadvantage between identical DCTCP flows. This raised the suspicion that at least part of the problem might be in the AQM, not the CCA. The working assumption is that the AQM marking might be biased against flows arriving in smaller bursts.

Further investigation~\cite{Briscoe17b:sigqdyn_TR} identified the likely cause: the use of sojourn time as the metric for queuing delay in the DCTCP AQM. The likely cause is not the use of queuing delay itself, but using sojourn time to measure it. Although the sojourn time of a particular packet is measured at dequeue, it measures the time to service the queue that was in \emph{front} of the packet when it arrived. Sojourn-based marking therefore indicates how other traffic affects each packet, not how each packet affects other traffic. For that, one ought to measure the queue \emph{behind} each packet when it departs.

This distinction becomes critical when flows of different burstiness are mixed. Consider a thought experiment where the goal is two flows of equal rate. Then, if one arrives in bigger bursts, they will arrive less often. The bigger the burst arriving at an empty queue, the longer some of it remains in the queue. So packets in frequent small bursts will be much more likely to sit behind, not in front of less frequent but larger bursts.

Then, marking based on sojourn time will be much more likley to mark the packets arriving in small bursts, because they have a larger queue in front of them. The report on `Rapid Signalling of Queue Dynamics'~\cite{Briscoe17b:sigqdyn_TR} calls this the `Blame Shifting Problem', because a sojourn-based AQM shifts much of the blame for queuing away from the bursty traffic that is largely responsible and onto smoother traffic that is largely blameless. The report proposes various ways of measuring the backlog as a packet departs then estimating the likely worst-case queue delay of the tail packet behind it. It uses the phrase `Expected Service Time' or EST as a general name for any of these techniques.

The blame shifting problem of sojourn-based marking is likely to have a number of negative connotations. In this section we focus on whether it is the cause of the latecomer disadvantage, where an established Linux DCTCP flow has larger burst size than an identical Linux DCTCP flow trying to push in. And we determine whether EST-based marking can eliminate or at least mitigate the latecomer disadvantage.

For this, we implement an AQM algorithm that uses
EST instead of sojourn time. We then test two identical DCTCP flows arriving 5\,s apart over this algorithm and compare the result with sojourn-based marking. We also try the different variants of the DCTCP EWMA in each case.

For the readers convenience we have renamed the DCTCP variants used in this
section:
\begin{description}
	\item[LoRes Toggle EWMA] (DCTCP-PS10Tu) the current default
	DCTCP implementation;
	\item[LoRes EWMA] (DCTCP-PS10tu), as default except no toggle of alpha to 0;
	\item[HiRes EWMA] (DCTCP-PS20tU), no toggle double precision alpha (1 in \(2^{20}\)) and EWMA stored upscaled. 
\end{description}
In
addition these variants all have a version with TSO disabled indicated by
"(wo TSO)" after its name.

Each of the variants are prefixed by either "EST" or "SOJ", meaning:
\begin{description}
	\item[EST]: the proposed expected service time metric;
	\item[SOJ]: the sojourn time metric
\end{description}
In all of the experiments the AQM uses a step marking
threshold at 2\,ms, and the marking is done on dequeue.

We also test whether the size of the max burst delay affects the latecomer disadvantage. For this we use TCP Prague, which already implements a parameter to set the max burst delay, whereas DCTCP inherits the burst duration implemented for BBR, which was hard-coded. Again, for reader convenience, we have given the resulting TCP Prague variants
more understandable names, as follows:
variants:\begin{description}
	\item[prague-1ms] has parameter value 10, meaning \((1/2^{10})\)\,s, which is just under
	1\,ms;
	\item[prague-250us] (default) has parameter value 12 or just under \(250\,\mu\)s;
	\item[prague-No	burst] has parameter value 20 or just under \(1\,\mu\)s, which effectively disables TSO over the range of rates in our investigation.
\end{description}

\subsection{DCTCP}
\begin{figure}
  \centering
  \includegraphics[width=.8\linewidth]{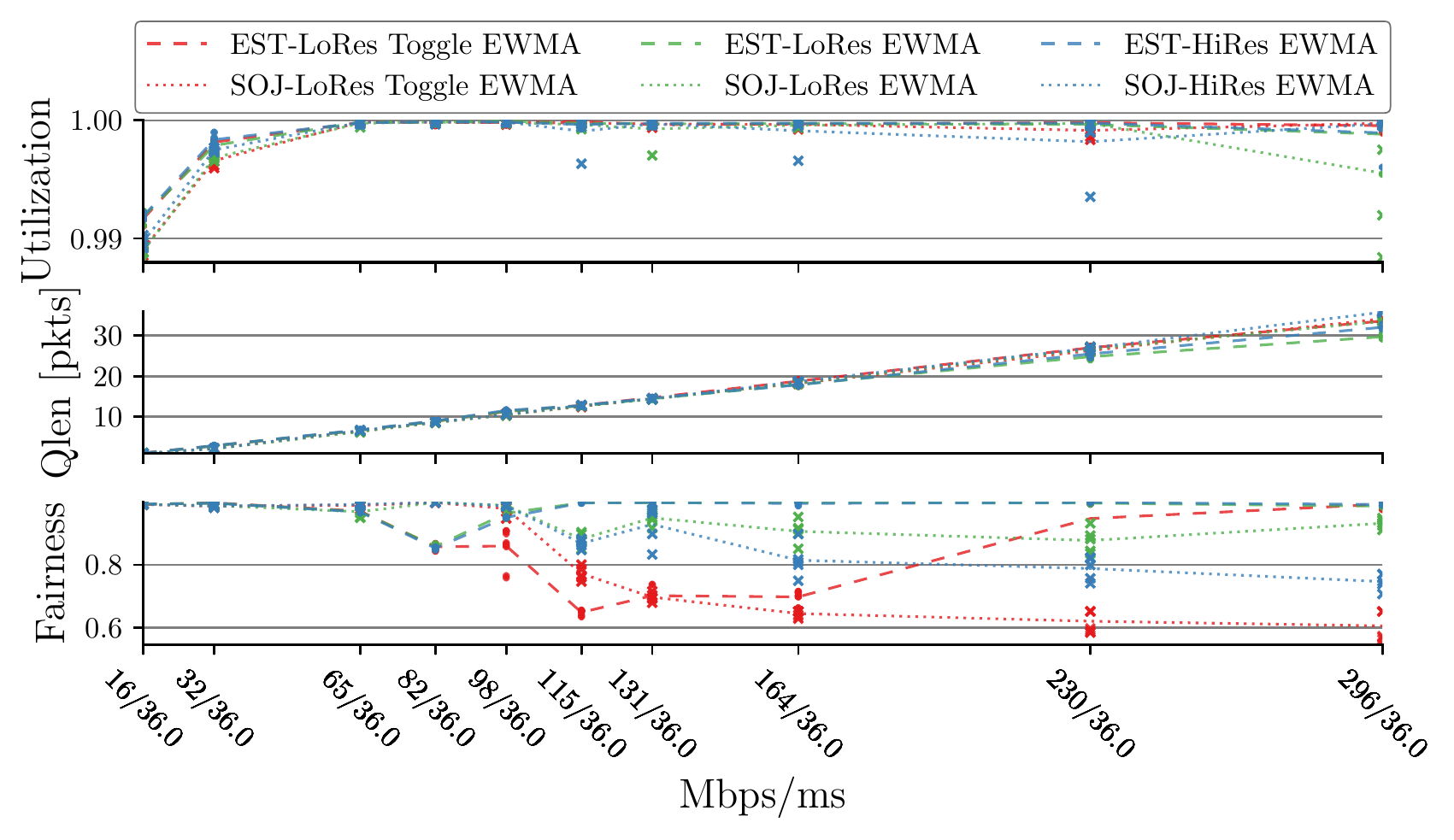}
  \caption{Over a range of link rates, EST seems to all-but eliminate the latecomer disadvantage as long as alpha is not
    toggled to 0 (the two non-toggling EST cases are hard to see, because they overlap each other along the 1.0 fairness index ceiling). The plots show average Jain's fairness index as well as utilization and queue length of two identical flows started 5\,s apart, for all six combinations of AQM metric and DCTCP variant.\\
    Capacity: varied; RTT: 36\,ms; AQM: step at 2\,ms.}
  \label{fig:toggle:tso_util_queue_fairness_vary_rtt_compare_capacity_dctcp}
\end{figure}
\begin{figure}
	\centering
	\includegraphics[width=.8\linewidth]{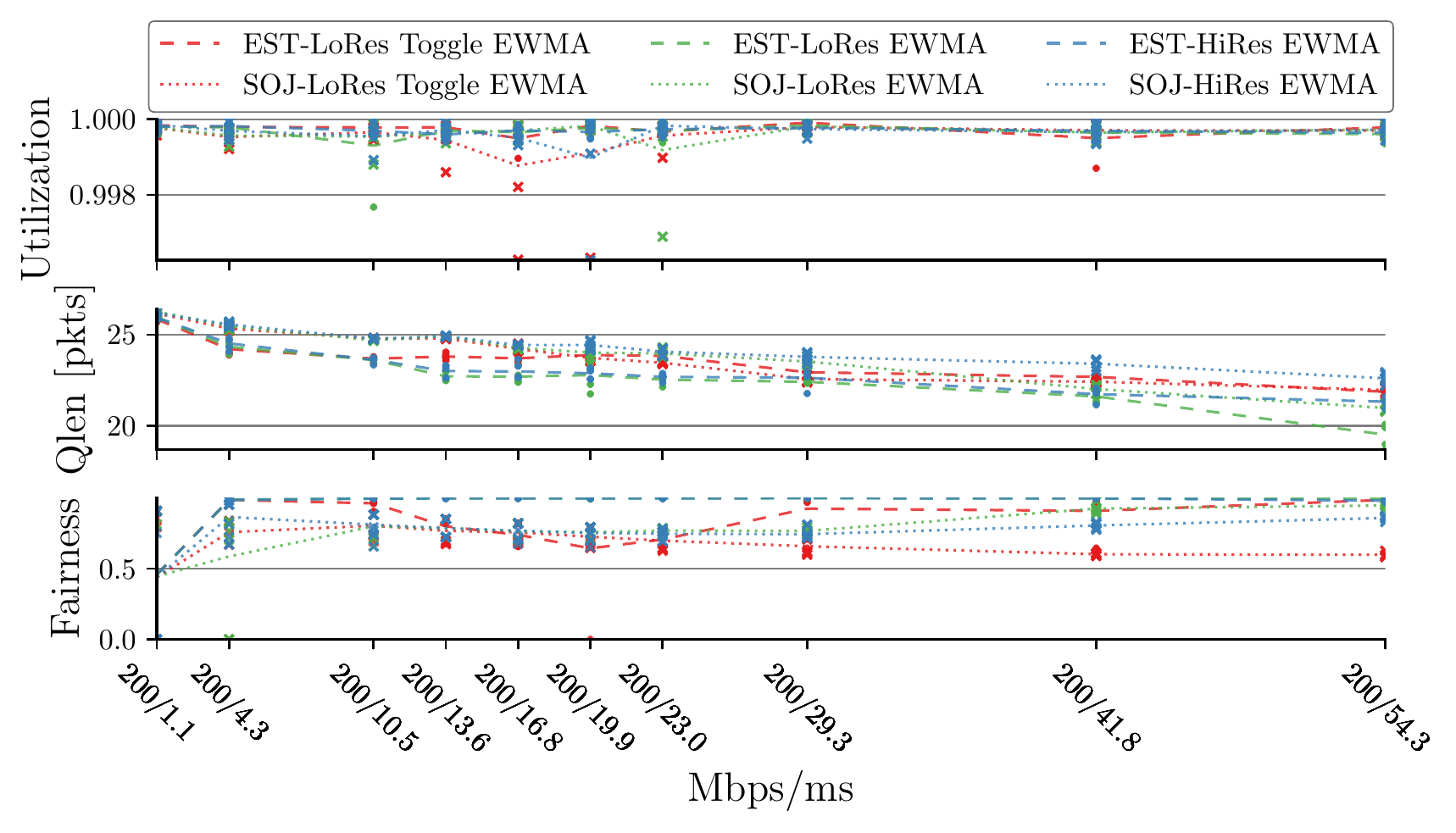}
	\caption{Over a range of RTTs, EST seems to eliminate the latecomer disadvantage as long as alpha is not
		toggled to 0 (the two non-toggling EST cases are again hard to see along the 1.0 fairness index ceiling). The metrics and six combinations of set-up are the same as in \Cref{fig:toggle:tso_util_queue_fairness_vary_rtt_compare_capacity_dctcp}\\
		Capacity: varied; RTT: 36\,ms; AQM: step at 2\,ms.}
	\label{fig:toggle:tso_util_queue_fairness_vary_rtt_compare_rtt_dctcp}
\end{figure}

The bottom rows of \Cref{fig:toggle:tso_util_queue_fairness_vary_rtt_compare_rtt_dctcp,fig:toggle:tso_util_queue_fairness_vary_rtt_compare_capacity_dctcp} show
average fairness between a pair of identical competing flows started 5\,s apart. All six combinations of CCA and AQM variants are tested, but each pair of flows within one run uses the same variant. 

EST seems to eliminate the latecomer disadvantage as long
as alpha is not toggled to 0.\footnote{Except for the uncharacteristic result at 82\,Mb/s, which might be due to some unfortunate synchronization effect.} In contrast, the sojourn-based AQM seems to cause
fairness issues in most scenarios. When capacity is low, the difference in
TSO burst size becomes insignificant. So, whether EST or SOJ is used becomes moot.

Utilization and queue length are also plotted as sanity checks.

The flows are both allowed to reach steady state by waiting 60\,s before measurements start. For each run, the metrics are calculated every two base round trips then averaged over a duration of 120\,s. For each combination of AQM and CCA, five runs are plotted as points to show the spread, then the line is plotted through the average of the five runs. 

The metric used for fairness is Jain's Fairness Index rather than the simple ratio between the flow rates. This is because the ratio between one rate and another might toggle between more than 1 and less than 1, so averaging would hide a toggling inequality. In contrast, Jain's index measures how unequal flow rates are irrespective of which is greater. So it should be robust to averaging over inversions.

\subsection{TCP Prague}
\begin{figure}
  \centering
  \vspace{-12pt}
  \includegraphics[width=.8\linewidth]{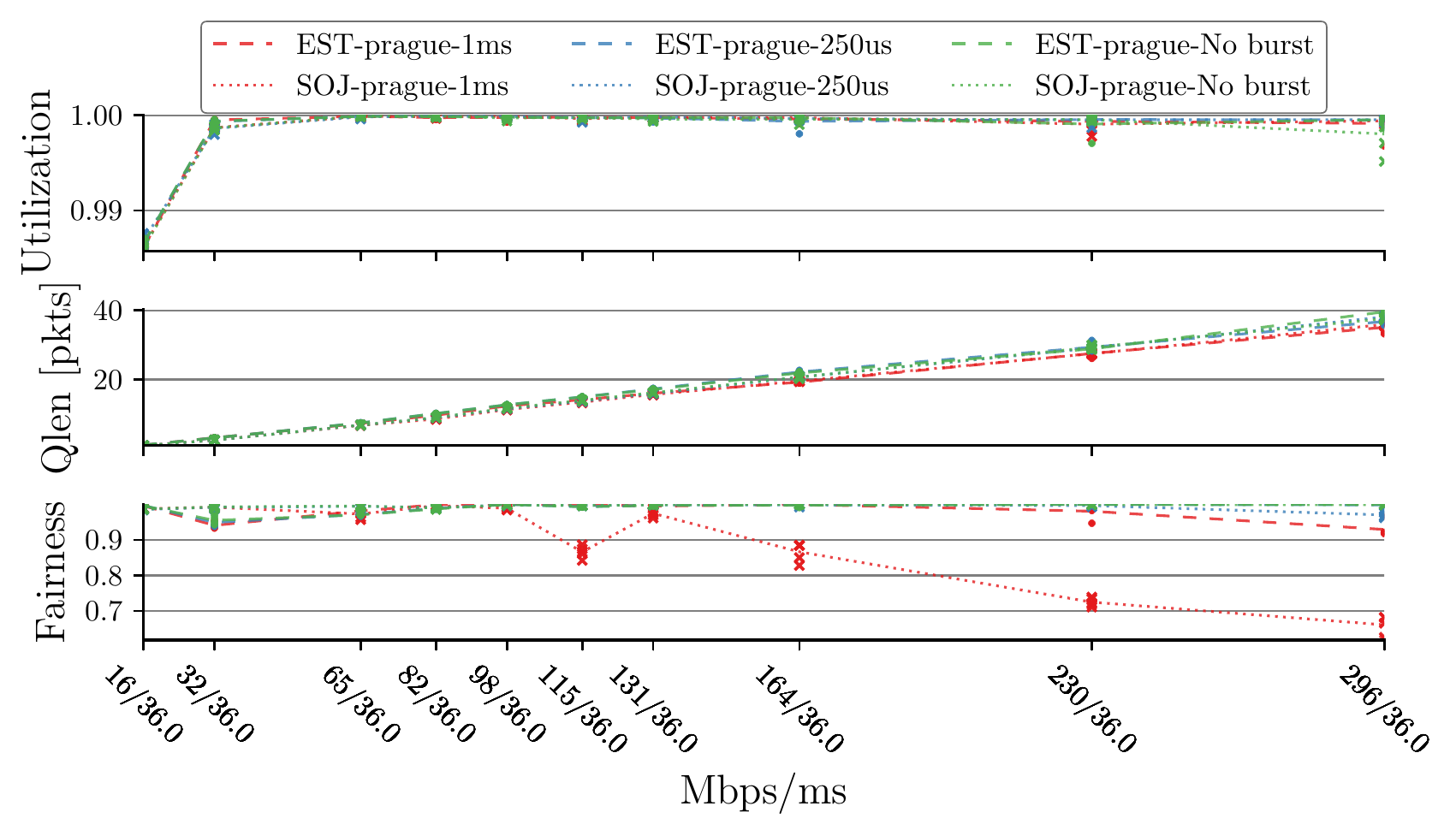}
  \vspace{-12pt}
  \caption{EST marking seems to fix the latecomer disadvantage over a sweep of link-rates, particularly in combination with reduced burst duration (the fairness plots for EST-prague-250 and TSO disabled all overlap along the 1.0 fairness ceiling). The plots show average utilization, queue length and fairness for the two delay metrics at the AQM (EST and sojourn time) and three different TSO burst sizes at the sender.\\
    Capacity: varied; RTT: 36\,ms; AQM: step at 2\,ms.}
  \label{fig:toggle:tso_util_queue_fairness_vary_rtt_compare_capacity_prague}
\end{figure}
\begin{figure}
  \centering
  \includegraphics[width=.8\linewidth]{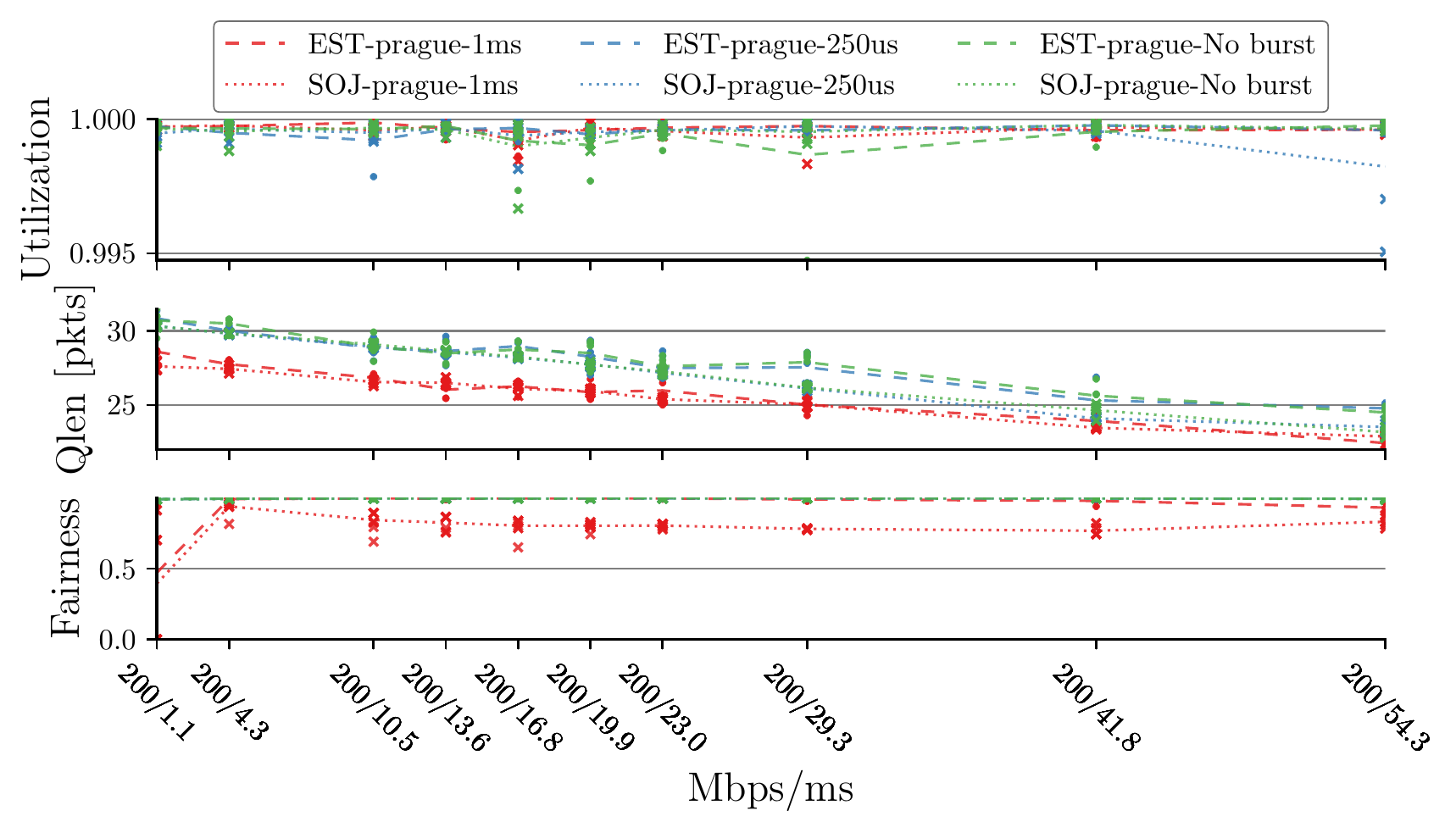}
  \vspace{-12pt}
  \caption{EST marking and lower burst size also seem to fix the latecomer disadvantage over a sweep of RTTs (the fairness plots for EST-prague-250 and for low or no TSO all overlap along the 1.0 fairness ceiling). The metrics and six combinations of set-up are the same as in \Cref{fig:toggle:tso_util_queue_fairness_vary_rtt_compare_capacity_prague}\\
    Capacity: 200\,Mb/s; RTT: varied; AQM: step at 2\,ms.}
  \label{fig:toggle:tso_util_queue_fairness_vary_rtt_compare_rtt_prague}
\end{figure}
\Cref{fig:toggle:tso_util_queue_fairness_vary_rtt_compare_rtt_prague,fig:toggle:tso_util_queue_fairness_vary_rtt_compare_capacity_prague} shows
average utilization, queue length and fairness (jains) between two competing flows with
different CCA and AQM variants. This experiment was run to see if the choice of a
250us burst size in TCP Prague eliminates the fairness problem for TCP
Prague. It is clear that having a burst of 250us or no burst at all eliminates
or reduce the fairness issue. The figures also show that TCP Prague would have
had the same fairness issue if it had the same burst size as DCTCP (1\,ms).

\begin{figure}
  \centering
  \includegraphics[width=.8\linewidth]{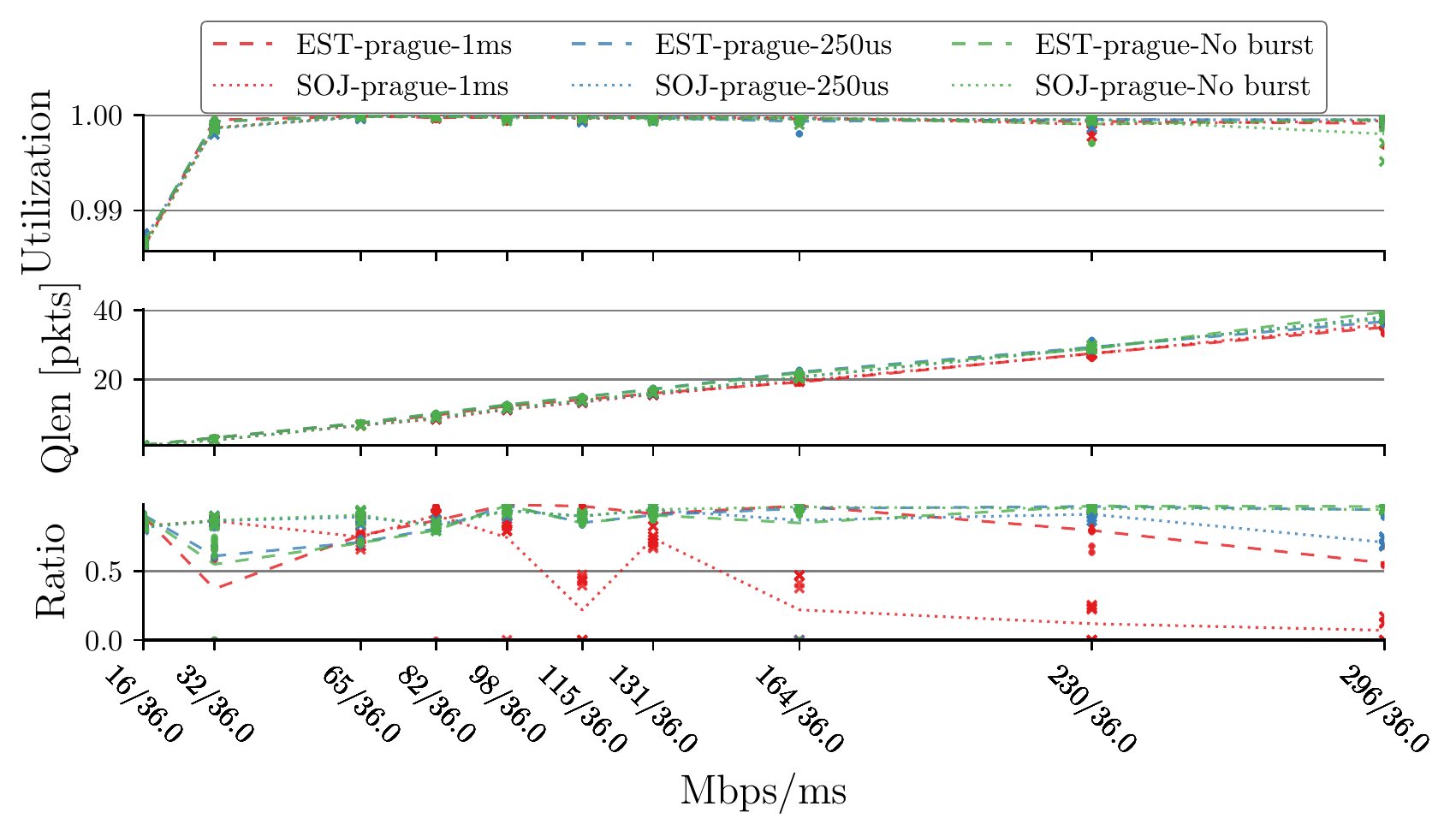}
  \caption{The same results as in \Cref{fig:toggle:tso_util_queue_fairness_vary_rtt_compare_capacity_prague} showing that smaller burst size seems to fix the latecomer disadvantage, except fairness is shown as a simple rate ratio.}
  \label{fig:toggle:tso_util_queue_fairness_vary_rtt_compare_capacity_prague_ratio}
\end{figure}
\begin{figure}
  \centering
  \includegraphics[width=.8\linewidth]{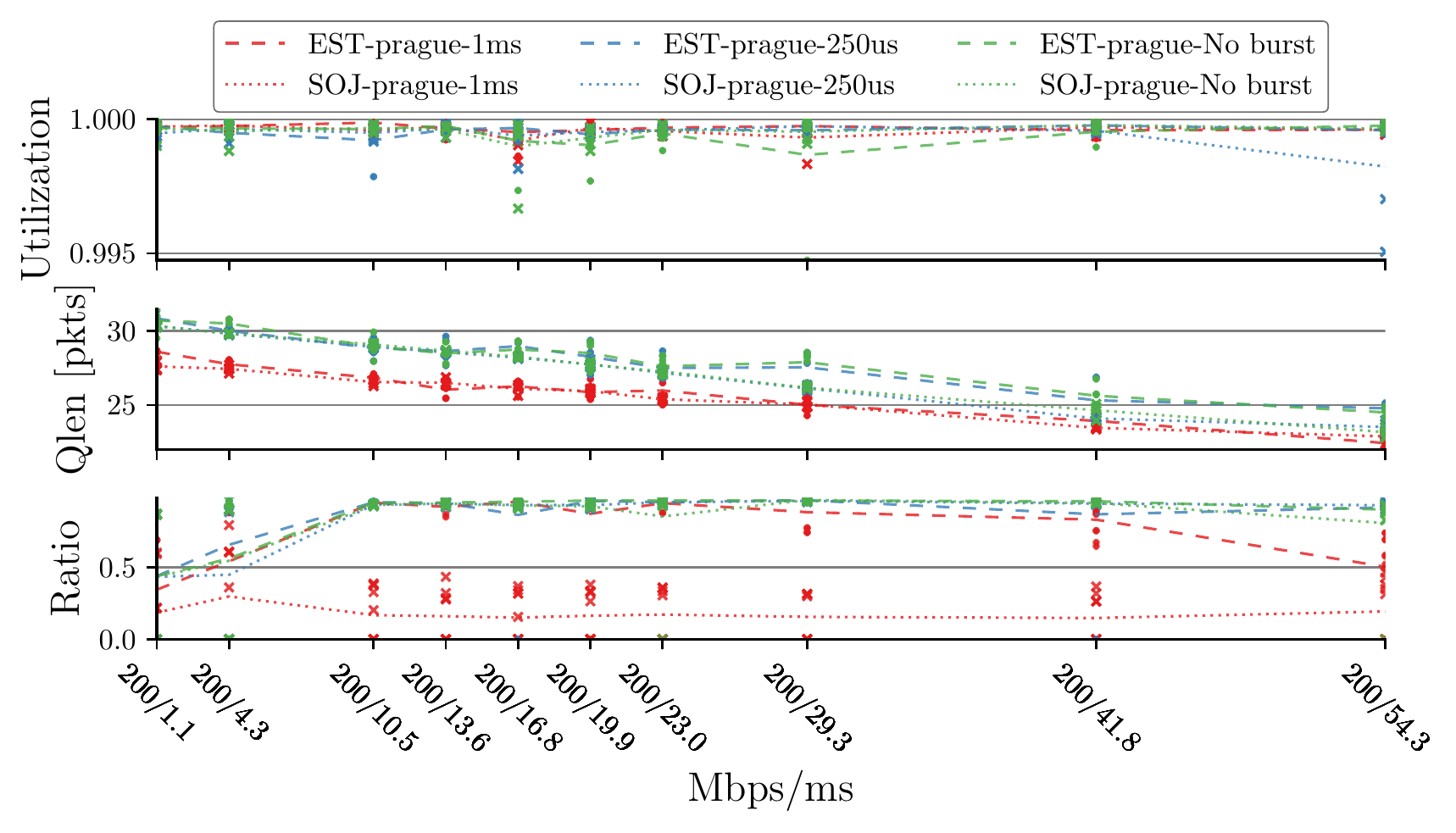}
  \caption{The same results as in \Cref{fig:toggle:tso_util_queue_fairness_vary_rtt_compare_rtt_prague} showing that smaller burst size seems to fix the latecomer disadvantage, except fairness is shown as a simple rate ratio.}
  \label{fig:toggle:tso_util_queue_fairness_vary_rtt_compare_rtt_prague_ratio}
\end{figure}
\Cref{fig:toggle:tso_util_queue_fairness_vary_rtt_compare_rtt_prague_ratio,fig:toggle:tso_util_queue_fairness_vary_rtt_compare_capacity_prague_ratio} show the same experiments as \Cref{fig:toggle:tso_util_queue_fairness_vary_rtt_compare_capacity_prague,fig:toggle:tso_util_queue_fairness_vary_rtt_compare_rtt_prague}, except the fairness metric is the averaged ratio between the two flow rates (using a geometric mean, but otherwise the same averaging approach). 

Previously, use of Jain's Fairness Index was justified on the grounds that merely averaging a rate ratio can cancel itself out in cases where a rate imbalance remains relatively constant but toggles between two flows. Here, we first check that there has been no toggling before showing the simple rate ratio, because it is more numerically meaningful as a figure of merit. For instance, when Jain's Index is 1/2 for two flow (or 1/3 for three), it is not obvious that this means one flow is causing the others to starve.

\clearpage
\section{Conclusions \& Recommendations}
\label{sec:tail-pieces}

\subsection{Conclusions}
\label{sec:conclusions}
We have shown that fixing the implementation of PRR in Linux by aligning the
code with RFC~6937 leads to proper cwnd adjustment.

However, fixing this bug in PRR unmasks a latecomer disadvantage
problem, where two long-running flows can get stuck in a `local equilibrium' with 
unequal rates. 

It is possible to mitigate the latecomer disadvantage in some cases 
by setting a lower floor to the 
average of ECN marking that DCTCP maintains. 
This probably works by making the queue variations large enough to jolt the flows out of their local equilibrium so they drop into the global equilibrium where rates are equal.
However, the floor makes DCTCP respond with a window decrease that is fixed at the set minimum rather than based on the extent of congestion. So, it effectively turns DCTCP into a classical congestion control, losing all the advantages of a scalable control, such as a short invariant recovery time between congestion signals and very low queue delay variation.

In preference to losing the benefits of scalable congestion control by just masking the latecomer disadvantage, 
we believe we have tracked down the root cause. It is actually a cross-layer
problem, due to an interaction between the TSO burst sizing in Linux and the instantaneous
(stateless) AQMs used with DCTCP:
\begin{itemize}
	\item Linux scales the size of TSO bursts in proportion to cwnd. So, a flow already using the capacity sends in larger bursts, while a latecomer, sending more slowly while it tries to push in, keeps its bursts smaller. 
	\item We have found that an 
	instantaneous AQM that applies ECN marking to packets based on their own sojourn 
	time perversely biases marking in favour of larger bursts and punishes 
	smoother traffic, which we call `blame shifting'. 
\end{itemize}
When this TSO sizing interacts with this perverse AQM marking, it can be seen that the latecomer will experience proportionately more ECN marking, which explains why it gets stuck at a lower rate.

A separate report~\cite{Briscoe17b:sigqdyn_TR} explains this `blame-shifting' problem 
in more depth. It discusses 
criteria for ensuring that marking probability increases with burstiness and suggests 
various design ideas for attributing blame to the most bursty traffic. 
The present report assesses an implementation of one of those ideas and finds that it appears 
to remove the latecomer disadvantage.

We also find that, if a solution to the blame-shifting problem is not implemented in the AQM, then removing all the truncation and precision bugs from the EWMA actually makes the latecomer disadvantage a little worse in steady state (i.e.\ with just two long-running flows and no background traffic). This is because the biased congestion marking from the AQM is averaged and stored precisely by the CCAs, with minimal noise to jolt them out of their local equilibrium.

\subsection{Recommendations}
\label{sec:recommendations}
\begin{enumerate}
	\item Apply the patch recommended to fix the Linux PRR bug.\footnote{Actually, it was patched in May 2022 after we notified the maintainer with an early copy of this report~\cite{Cheng22:Linux_PRR_bugfix}} Strictly, this should be patched after the blame-shifting problem has been addressed. However, even though there will no longer be noise due to the PRR bug to mask the latecomer disadvantage, in many circumstances there is enough noise from background traffic to do the same job.
	\item In immediate AQMs for scalable traffic, address the `blame-shifting problem' of sojourn marking (our own research on alternatives is ongoing and will be reported in future versions of \cite{Briscoe17b:sigqdyn_TR});
	\item Investigate whether dithering additive increase or burst size jolts flows out of their local equilibrium and therefore mitigates the latecomer disadvantage, even when the AQM has not been fixed;
	\item Remove the 'toggle to zero' (line 123 in \cref{listing-dctcp-original}) and the blackholing of congestion feedback (lines 130 \& 131) from Linux DCTCP and any other Scalable CCAs (they are already removed from Prague). A floor of 16/1024 should not be left in place of the toggle, otherwise incremental deployment of a full solution will be hard in future (see earlier footnote\textsuperscript{\ref{note:incr_deploy_DCTCP}});
\end{enumerate}

\subsection{Further Work}
\label{sec:further_work}
\begin{itemize}
	\item Completion of the evaluation of remedies to the `blame-shifting problem' with sojourn marking, as already highlighted above~\cite{Briscoe17b:sigqdyn_TR}
	\item Evaluation of further improvements to scalable CCAs for instance:
	\begin{itemize}
		\item Investigation of dithering additive increase or burst-size (e.g.\ by carrying the remainder after rounding): i) in order to jolt flows out of any undesirable local equilibrium (as above) and ii) in order to emulate the smooth marking behaviour of a ramp AQM, when there is a step AQM at the bottleneck;
		\item Evaluation of continuing additive increase during CWR state;
		\item Evaluation of per-ACK EWMA and per-ACK congestion reduction to remove the clock machinery lag in scalable congestion controls (see \cite{Briscoe22a:Prague_clock_lag}).
	\end{itemize}
\end{itemize}

\clearpage
    \bibliography{main} 
    \bibliographystyle{ACM-Reference-Format}

\clearpage
\appendix

\section{DCTCP and its Implementation in Linux}
\label{appendix:dctcp-algo-and-implementation}
DCTCP is an algorithm that can achieve high throughput and low latency at the same
time. It is able to do so because of two changes, one to the end-system and one
to the network. In the network long buffered tail drop queues are replaced by
shallow buffered queues that mark packets with explicit congestion experience
(ECN) marks in the IP-header. The extent of these marks is then used to reduce
the congestion window proportionally to the extent of the marks in the
sender. The change to the sender reduce congestion window oscillation
sufficiently to keep high utilization.

DCTCP is designed to work with an AQM that uses a step threshold to decide
whether or not to mark each packet. The queue occupancy at which the threshold
is set affects latency and utilization. If the threshold is set extremely 
shallow it can lead to under-utilization. Setting the threshold too high cause
latency to increase. There is no single ideal configuration. One can sacrifice some
utilization to get lower latency. The ideal configuration for keeping full
utilization is to have the
oscillation in the queue length reach the minimum number of packets required to
keep full utilization. This way the latency is minimized and utilization is
maximized.

Despite having been designed to work with a step AQM, DCTCP also works with
probabilistic AQMs. A probabilistic AQM can provide more a fine grained marking
proportion to the sender that can be used to reduce oscillation and thereby
allow for reduction in queueing delay without reducing utilization.

DCTCP maintains an exponential weighted moving average (EWMA) of the proportion of bytes
carrying explicit congestion notification. The average proportion is called
alpha, and is denoted by $\alpha$ below. Roughly every round-trip time $\alpha$ is updated
according to the following equation. The start and completion of a round-trip
time is stored and detected using sequence numbers.

\begin{equation}
  \alpha = (1-g) * \alpha + g * F
\end{equation}
\noindent where
\begin{description}
\item $g$ is the EWMA factor. The default value in DCTCP is $\frac{1}{16} \left(0.0625\right)$.
\item $F$ is the measured \textbf{F}raction of bytes carrying ECN marks of the
  delivered packets over the previous round-trip time. $F \in [0,1]$
\end{description}

\noindent
Every time an ack carrying ECN is received, and there has been no reaction to ECN
the past round-trip time, the congestion window, $W$, is reduced by $\alpha * W
/ 2$.
\begin{equation}
  W \gets W - \frac{\alpha * W}{2}
\end{equation}
The congestion window is reduced at most once per round-trip time. On each ack the
congestion window is increased by $1/W$. This is more commonly known as
additive increase. The Linux implementation implements
this increase as a separate count variable that is used to increase $W$ by 1
once the count grows big enough. The aggregate change in congestion window in a round with a
reaction to congestion is thus

\begin{equation}
  W \gets W + 1 - \frac{\alpha * W}{2}
\end{equation}

A DCTCP flow is therefore in equilibrium when $\alpha = 2/W$ resulting in a net change of
0 over a round-trip time with a reaction.

\Cref{listing-dctcp-original} on \pageref{listing-dctcp-original} shows relevant parts of the current (Sep 2021) DCTCP implementation
in Linux. In line 108 the congestion window is
reduced based on the current value of $\alpha$ and \(W\). Because kernel code must not
use floating point operations dctcp\_alpha is scaled by a factor of 1024, or 10
bit-shifts to the left. That
is why the result of the multiplication is shifted by 10 to scale it down and
shifted by 1 to divide it by 2.

Line 123 executes the first part of the EWMA, where $\alpha$
is subject to a decay. We discuss why this particular line is
problematic for the scalability of DCTCP in \cref{sec:toggling-problem}.

The calculation of the fraction of ECN-marked bytes, $F$, occurs on line 130 and 131. The number of bytes
marked with congestion experience (CE) is scaled and divided by the total number
of bytes delivered. The problem with those lines black-holing congestion feedback is also described in \cref{sec:toggling-problem}. Finally, at line 139 alpha is updated with its new value.

\subsection{Scalability}
\label{sec:scalablity}
DCTCP is called a scalable CCA because its steady-state sawtooth
does not grow as capacity increases. In fact, if the proportion of marked bytes
stored in $\alpha$ exactly matches $2/W$ there is theoretically no oscillation
at all because the reduction and increase each round-trip time cancel each
other. They key to scalability is that the oscillation can be kept constant when
capacity increases. An increase in capacity requires a increase in $W$ to keep
full utilization, but as long as $\alpha * W$ stays the same the oscillation
will stay the same. We have that

\begin{equation}
  \alpha = \frac{2}{W}
\end{equation}

Alpha must decrease to keep the
algorithm in equilibrium as $W$ increases with increasing capacity. If $alpha$
fails to decrease the oscillation grows and fails to provide theoretical
scalability. This puts a requirement on the representation of $\alpha$ in any
implementation of a scalable CCA.
It must be able to represent very small values of $\alpha$.

In practice it is impossible to keep alpha at the exact equilibrium
value, and the equilibrium value is not static. Representability is not enough
to keep DCTCP scalable in practice, the
adjustments on alpha needs to be small as well. As alpha decreases an
implementation should make smaller and smaller adjustments to alpha so that
noise does not result in unwanted oscillation.

Even with a perfect representation of alpha in the CCA the
marking proportion from the network has to decrease as capacity increases to
keep window oscillation the same. Theoretically, the proportion can decrease because, by definition, there are more packets in a round as the window increases. So, as long as the same average number of packets per round are marked (specifically an average of 2 marks per RTT), the proportion will decrease. However, this is hard to achieve with a step threshold because
it tends to induce an on-off pattern of marking with whole rounds marked or not (see the full discussion of this in \cref{sec:toggle_eval_ramp} about using a ramp for marking).

\section{Machine configuration}
\label{appendix:machine-config}

We have run experiments on two separate testbeds. The difference between them is
mainly the hardware.

\subsection{Wide area network testbed}
The testbed consist of three machines and we will call them client, server and
aqm.

Both the client and the server runs Ubuntu 18.04.4 with a 5.9.0-X Linux
kernel. X represent a specific version. The aqm runs Ubuntu 16.04.5 with a
5.4.0-rc3 Linux kernel.
Each machine has a 1 Gigabit Intel I350 Gigabit network card.
The driver is igb version 5.9.0-X. The firmware is version 1.63.
The client and the server has a 2-port version of the card, while the aqm has
a 4-port version of the card.

The machines are connected in serial and through a switch. The switch makes it
possible to have management traffic between the machines without interfering
with the experimental traffic.

Delay and rate limiting is added by tc-netem and tc-htb respectively on the aqm
machine. To avoid timer related issues that we have previously experienced using
tc-htb and tc-netem we set the cpu scaling governor to performance and the
highest C-state to 0 on every cpu on the aqm.

Address resolution protocols, arp, tables are static. This eliminates possible
stops in transmission due to address resolution.

The following interface capabilities are turned off on every interface using
ethtool: gso, tro, gro, tx-gso-partial, autoneg, sg. In addition rx-usecs and
tx-usecs are
set to 0 on all the intefaces.

\noindent
The following sysctl settings are set to 1: tcp\_no\_metrics\_save,
tcp\_low\_latency, tcp\_tw\_reuse, tcp\_ecn.

\noindent
The following sysctl settings are set to 0: tcp\_autocorking, tcp\_fastopen,
tcp\_ecn\_fallback.

\noindent
Memory limits are set to 8388608 through the following sysctl settings:
rmem\_max, wmem\_max, rmem\_default, wmem\_default, tcp\_rmem, tcp\_wmem,
tcp\_mem

\subsection{Data centre testbed}

The testbed we used for data centre like experiments consist of three machines
connected in serial, called client, server and
aqm. Management traffic avoids experimental traffic by going
through a separate network. Delay and rate limiting is added by tc-netem and
tc-htb respectively on the aqm machine.

All three machines are running Debian stretch (9.2). The aqm machine runs Linux
4.9.0, while the client and server runs 5.9.0-X.

All three machines have a 10 Gigabit X550T Intel Ethernet network card. The
driver is ixgbe version 5.9.0-X. The firmware is presented as 0x80000492 in
lshw.

\noindent
The following sysctl settings are set to 1: tcp\_no\_metrics\_save,
tcp\_low\_latency, tcp\_tw\_reuse, tcp\_ecn.

\noindent
The following sysctl settings are set to 0: tcp\_fastopen,
tcp\_ecn\_fallback.

\noindent
Memory limits are set to 8388608 through the following sysctl settings:
rmem\_max, wmem\_max, rmem\_default, wmem\_default, tcp\_rmem, tcp\_wmem,
tcp\_mem

\section{Code details}
\label{appendix:code_variants}
The Prague implementation used has commit ID
3cc3851880a1\footnote{\url{https://github.com/L4STeam/linux/commit/3cc3851880a1b8fac49d56ed1441deef2844d405}}.

\subsection{Diffs for DCTCP implementations}\label{sec:dctcp-diffs}
	\begin{table}[h]
	\newcommand{\difhref}[2]{\href{https://bitbucket.org/joakimmisund/phd-shared-documents/src/master/papers/prr-bug/paper/diffs/tcp_dctcp_#2_diff.txt}{#1 (#2)}}
	\begin{tabular}{l c c c c c}
		Name & PRR & TSO/GSO & Alpha precision (bits) & Alpha floored & Upscaled\\
		\toprule 
		\difhref{DCTCP-PS10Tu}{a} & yes & yes & 10 & yes & no \\
		\difhref{DCTCP-pS10Tu}{b}  & no  & yes & 10 & yes & no \\
		\difhref{DCTCP-Ps10Tu}{c}  & yes & no  & 10 & yes & no \\
		\difhref{DCTCP-PS20tU}{d}  & yes & yes & 20 & no  & yes \\
		\difhref{DCTCP-pS20tU}{e}  & no  & yes & 20 & no  & yes \\
		\difhref{DCTCP-PS10tu}{f}  & yes & yes & 10 & no  & no \\
		\difhref{DCTCP-ps20tU}{i}  & no  & no  & 20 & no  & yes \\
		\difhref{DCTCP-Ps10tu}{j}  & yes & no  & 10 & no  & no \\
		\difhref{DCTCP-Ps20tU}{k}  & yes & no  & 20 & no  & yes \\
		\difhref{DCTCP-pS10tu}{l}  & no  & yes & 10 & no  & no \\
		\difhref{DCTCP-ps10tu}{m}  & no  & no  & 10 & no  & yes \\
		\difhref{DCTCP-ps10Tu}{n}  & no  & no  & 10 & yes & no \\
		\difhref{DCTCP-Ps10tU}{o}  & yes & no  & 10 & no  & yes
	\end{tabular}
	\caption{Tested DCTCP variants and their differences. The names of each variant provide clickable links to the diff of each wrt.\ the DCTCP code (as of 1 Sep 2021).}
	\label{tab:dctcp-variants}
\end{table}

\end{document}